\documentclass[review]{elsarticle}
\usepackage[english]{babel}
\usepackage{booktabs}
\usepackage{multirow}
\usepackage{lineno,hyperref}
\usepackage{amsfonts}
\usepackage{dsfont}
\modulolinenumbers[5]
\usepackage[utf8]{inputenc} 
\usepackage{mathtools}
\usepackage{amssymb}
\usepackage[dvipsnames]{xcolor}
\usepackage{lscape}
 \usepackage{tabu}
\usepackage{verbatim}
\usepackage{algorithm}
\usepackage{algorithmic}
\usepackage{calrsfs}
\DeclareMathAlphabet{\pazocal}{OMS}{zplm}{m}{n}

\usepackage{setspace}
\usepackage{multirow}
\journal{Journal of \LaTeX\ Templates}

\journal{Information Sciences}

\begin{document}

\begin{frontmatter}

\title{Scalable Teacher Forcing Network for Semi-Supervised Large Scale Data Streams}

\tnotetext[mytitlenote]{Corresponding Author}

\author[mymainaddress]{Mahardhika Pratama*}

\author[myaddress]{Choiru Za'in}

\author[mythirdaddress]{Edwin Lughofer}
\author[mysecondaryaddress]{Eric Pardede}
\author[myfourthaddress]{Dwi A.P. Rahayu}

\address[mymainaddress]{School of Computer Science and Engineering, Nanyang Technological University, Singapore}
\address[myaddress]{Faculty of Information Technology, Monash University, Australia}
\address[mythirdaddress]{Department of Knowledge-Based Mathematical Systems, Johannes Kepler University, Linz, Austria}
\address[mysecondaryaddress]{Department of Computer Science and IT, La Trobe University, Australia}
\address[myfourthaddress]{Informatics Department, Faculty of Engineering, Universitas Muhammadiyah Malang (UMM), Indonesia}

\begin{abstract}
The large-scale data stream problem refers to high-speed information flow which cannot be processed in scalable manner under a traditional computing platform. This problem also imposes expensive labelling cost making the deployment of fully supervised algorithms unfeasible. On the other hand, the problem of semi-supervised large-scale data streams is little explored in the literature because most works are designed in the traditional single-node computing environments while also being fully supervised approaches. This paper offers  Weakly Supervised Scalable Teacher Forcing Network (WeScatterNet) to cope with the scarcity of labelled samples and the large-scale data streams simultaneously. WeScatterNet is crafted under distributed computing platform of Apache Spark with a data-free model fusion strategy for model compression after parallel computing stage. It features an open network structure to address the global and local drift problems while integrating a data augmentation, annotation and auto-correction ($DA^3$) method for handling partially labelled data streams. The performance of WeScatterNet is numerically evaluated in the six large-scale data stream problems with only $25\%$ label proportions. It shows highly competitive performance even if compared with fully supervised learners with $100\%$ label proportions.   
\end{abstract}


\end{frontmatter}


\section{Introduction}

\noindent\textbf{Background}: The problem of data streams \cite{Gama10} has attracted growing research interest because of its importance to handle current real world problems where data samples are collected continuously in never-ending and dynamic environments \cite{LughoferMouchaweh19}. This issue requires not only fast data processing with limited memory burden but also capability in handling distributional variations of data streams. The problem of data streams have been addressed using different approaches including rule-based approach \cite{CanoK19}, flexible decision tree \cite{VFDT}, ensemble methods \cite{PratamaPedryczLughofer18} \cite{SidhuBhatia15} \cite{JiangZhaoLu14} \cite{DingWangLiChaiWang17}, evolving neural networks \cite{Kasabov07} or evolving neuro-fuzzy systems \cite{SkrjancIglesiasLughoferGomide19} \cite{LughoferChapter14} and recently deep neural networks \cite{Aggarwal18,AshfahaniP19,PratamaZAO019}. Typically, data stream mining requires techniques in the field of drift detection and handling within reasonable space, time and memory complexities \cite{KhamassiMouchawehHammamiGhedira16} \cite{LughoferWeiglHeidlEitzingerRadauer16}. Nonetheless, the issue of data streams deserves in-depth study because of the increasing challenges of data analytics in practice such as limited access of ground truth and explosion of data volumes \cite{CanoK19}. The first one is known as the problem of semi-supervised learning while the second one is understood as the large-scale data stream learning problem. 

Large-scale data streams generate massive information flow which cannot be efficiently handled in a single computing node \cite{BifetHolmesKirkbyPfahringer10}. Unlike conventional big data problem where the size of information is constant and processed in the one-shot fashion \cite{Dean12} \cite{WuZhuWuDing14}, it refers to continuously growing information having to be processed in the single scan \cite{Gama10}. It calls for a distributed computing strategy being able to handle the never-ending information flow. Recently, there has been also a trend to perform the distributed computing task under the GPU in lieu of cloud-based infrastructure \cite{ErlPuttiniMahmood13,CanoK19}. The underlying challenge lies in the fact where the model's complexity grows exponentially if no proper complexity reduction step is implemented properly after the distributed computing phase. It should not suffer from accuracy loss. That is, the distributed computing structure should induce a similar or even improved accuracy, compared to the single node structure \cite{ZainAPLP20}. 

While massive information has to be executed in scalable fashion, annotation of large-scale data stream is prohibitive. This issue leads to an availability of only a small fraction of labelled samples (termed as scarcely labelled samples \cite{LughoferWeiglHeidlEitzingerRadauer16}), or, in the extreme case, labelled samples are only available during the warm-up phase, termed as \textbf{the infinite delay problem} \cite{SouzaSilvaBatistaGama15}. This issue prohibits the use of fully supervised learning algorithm and requires a label enrichment mechanism via pseudo-label generation \cite{incremental_hierarchical} to self-annotate unlabelled samples and the label augmentation approach to perturb labelled samples without changing their labels. The underlying challenge exists in the accumulation of own classification errors, which typically causes loss of generalization performance, because wrong labels are fed into the model update algorithm, thereby changing the shape of the decision boundary onto a wrong direction. 
On-line active learning is another possibility to decrease the annotation effort, but still it typically requires at least 10-20\% of the whole stream to be labelled in order to keep the model accuracy on a reasonable level \cite{Lughofer17} --- that is too much in large-scale streaming environments, where millions of samples are produced during a day or even an hour. Moreover, the online active learning cannot deal with partially labelled samples because true class labels of uncertain samples might not be obtained.

\noindent\textbf{Practical Scenario}: The problem of large-scale data streams in semi-supervised mode is exemplified in the monitoring problem of high-speed manufacturing processes. A manufacturing process runs 24/7 without interruption where data points are sampled from sensors in a high speed. Since data samples are generated in never-ending fashion, it creates streaming data environments in which a model has to be updated quickly while imposing low memory footprint. That is, a model is updated in a single scan where a data sample is discarded once used. In addition, another challenge exists in the dynamic and evolving characteristics of data streams \cite{Gama10} where data distributions rapidly change. This problem is portrayed in a manufacturing process operating in several conditions to produce a variety of products. Hence, a model is supposed to adapt to varying distributions on the fly without a retraining process from scratch. This problem also prohibits the use of a fixed model becoming outdated quickly. The large-scale data stream problem refers to a practical case in which the size of data streams is very large such that it is too demanding to be executed in a single node by using traditional data stream algorithms. This problem is evident in the high-speed manufacturing process producing a large amount of data samples in a short period. This problem calls for high-end computing infrastructure such as distributed computing paradigm under a cluster of computers \cite{za2019evolving} or GPUs \cite{CanoK19} to expedite model updates and thus innovative distributed learning algorithms. The underlying goal is to attain comparable accuracy as that in the single computing node while having significantly faster execution time than that in the single computing node. The large-scale data stream also brings a concern for the labelling cost because if not handled an operator has to annotate a massive number of data points in a short period. This issue leads to an urgent need to develop a semi-supervised algorithm which is capable of learning from few labelled samples. This case is evident in the monitoring problem of manufacturing cases because defect or worn cases are identified from manual inspection thus slowing down the process.  
\begin{figure}[htbp]
	\begin{centering}
		\includegraphics[width=13cm]{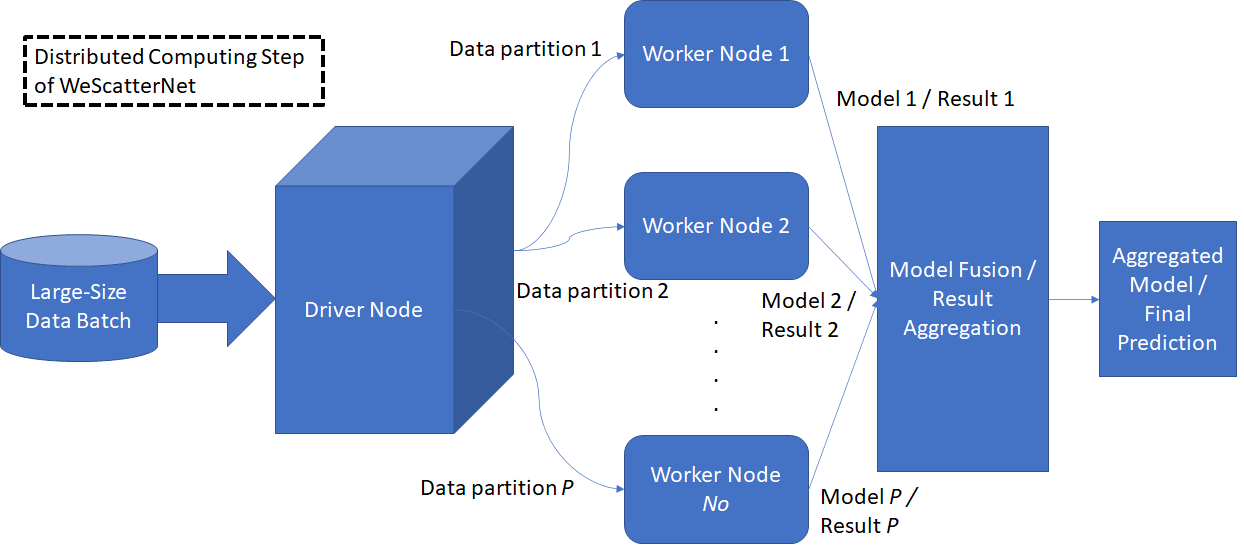}
		\par\end{centering}
	\caption{Distributed Computing Strategy of WeScatterNet: A large-size data batch is partitioned into \textit{P} data partitions by the driver node. \textit{P} data partitions are processed by $N_o$ worker nodes producing \textit{P} local results or base learners. The \textit{P} base models are consolidated into a single model and compressed to generate a compact model by the model fusion phase. Our work applies one executor per node where 6 worker nodes and 1 driver node are deployed.}
	\label{Distributed_Computing}
\end{figure}

\noindent\textbf{Our Contribution}: Weakly Supervised Scalable Teacher Forcing Network (WeScatterNet) is proposed in this paper. WeScatterNet addresses the large-scale data stream problem by means of data augmentation, annotation, and auto-correction ($DA^3$) while being executed in the distributed computing environment of Apache Spark to cope with massive information flow. WeScatterNet is a flexible ensemble network where both ensemble configuration and base classifier feature an open network structure to cope with varying data distributions. The base classifier can be dynamically added and pruned via the drift detection method. Furthermore, its base classifier characterizes a self-organizing network structure by growing and pruning mechanisms of fuzzy rules via the network significance (NS) method \cite{AshfahaniP19}. That is, it is capable of initiating its network structure from scratch or a predefined configuration while its fuzzy rules can be constructed or removed automatically with respect to varying data distributions. To meet parallel computing demands of large-scale streams in real-time, the base classifier is configured in the ensemble structure processed on distributed nodes, i.e. each node has its own neuro-fuzzy system (base learner) self-evolved and adapted over time on a specific data batch assigned to the corresponding computing node. Model fusion via the rule merging process is carried out afterward to compress the ensemble network into a single model. 

$DA^3$ is derived from the concept of MixMatch \cite{mixmatch} where it is equipped with a way to self-label unlabelled samples while labelled samples are enriched by perturbing them with controlled noise without changing their labels. In addition, an auto-correction mechanism is integrated to resolve the problem of wrong label representations due to the absence of ground truth. The conventional fuzzily weighted generalized recursive least square approach (FWGRLS) method is extended by incorporating the elastic weight configuration (EWC)-like regularization strategy \cite{kirkpatrick2016overcoming}. This is an improved $L_2-norm$ regularization method which takes into account the deflection due to the noisy pseudo-label and integrates a weight decay term to enhance generalization performance. This is different from \cite{PratamaAH19} where WeScatterNet is built upon the teacher-forcing principle and the fuzzy neural network paradigm. It is capable of explaining its operation via human-like linguistic fuzzy rules offering some sort of transparency.

The distributed computing strategy of Apache Spark is carried out in the continual fashion in which a data-free model fusion mechanism is executed after the distributed computing phase (one evolving model per node). It merges inconsequential rules in each of the base learners to those of high-quality rules thereby reducing the model's complexity without compromising predictive quality. It uses the concept of checking the distance and angle between the consequent hyper-planes of two rules, as well as the support compared to dominant rules. Fig. \ref{Distributed_Computing} pictorially illustrates the distributed computing strategy of WeScatterNet. Major contributions of this paper are summed up as follows: 
\begin{itemize}
\item WeScatterNet is proposed to handle a semi-supervised learning problem of large-scale data streams. WeScatterNet is developed as a self-organizing ensemble classifier where both base learner and ensemble structure possesses a self-evolving property to address concept drifts in the local level and in the global level. The novel aspect is seen in the integration of forgetting strategy in the NS method \cite{PratamaAH19} allowing to adjust the probability density function estimator in respect to the drift rate.   
\item Unlike existing ensemble classifiers for data streams, WeScatterNet is designed in the distributed computing strategy of Apache Spark to address the problem of large-scale data streams in scalable manner. That is, both the testing process and the training process are in parallel executed in the distributed computing nodes to speed up its computational time. Note that Apache spark here is put forward to cope with continuous information flow rather than a single-shot training process \cite{za2019evolving}. 
\item The data-free model fusion method is designed to compress an aggregated model after the distributed computing phase. It is capable of compressing the model's complexity during the distributed computing step without loss of accuracy. This strategy can be perceived as an improvement of the model fusion strategy in \cite{ZainAPLP20} where an online model selection phase is integrated. This strategy selects the best number of rules for a compressed model to retain predictive accuracy. 
\item Partially labelled data streams are handled by the $DA^3$ method performing the label enrichment and regularization mechanisms, The new aspect of $DA^3$ method lies in the new regularization strategy in the FWGRLS method to prevent the accumulation of errors as a result of noisy pseudo-labels. This regularization strategy is inspired by the EWC method \cite{kirkpatrick2016overcoming} originally devised to prevent the catastrophic forgetting problem of continual learning. Our approach tackles loss of generalization power due to noisy pseudo label where important rules are regularized such that they do not move too far from their optimal locations when updated by noisy pseudo labels. 
\item Our codes, data and raw numerical results are made publicly available in \href{https://github.com/ContinualAL/WeScatterNet.git}{\textit{\textbf{WeScatterNetCodeLink}}} to allow convenient reproduction of our numerical results and further study. 
\end{itemize}

Numerical results over six large-scale data streams coupled with the ablation study and the study of different class proportions have demonstrated the advantage of the proposed approach where it is capable of producing competitive performance using only $25\%$ label proportions compared to its fully supervised competitors having $100\%$ access of true class labels. The rest of this paper is structured as follows: Section 2 outlines related works; Section 3 discusses the problem formulation of semi-supervised data streams in the large-scale environments; Section 4 outlines the learning procedure of WeScatterNet; Section 5 elaborates our numerical study; Some concluding remarks are drawn in the last section of this paper. 

\section{Related Works}
The concept of evolving fuzzy system (EFS) is developed as a way to cope with the data stream problem where the key idea lies in the combination of parameter learning and structural learning under a single training phase. It enables fuzzy rules to be automatically constructed on the fly while performing a single-pass model update. It makes use of the local property of fuzzy rule where an online clustering technique can be benefited for automatic fuzzy rule generation while the rule consequent is updated using a local learning technique via the fuzzily weighted recursive least square (FWRLS) method. EFS does not utilize the evolutionary computing techniques for fuzzy rule construction \cite{Melin1,Melin2,Melin3}. This research area has started in the early 2000 where a pioneering work is proposed in \cite{eR} introducing the concept of incremental unsupervised learning. Another early work is proposed in \cite{DENFIS} using the evolving clustering method (ECM) for online identification of fuzzy rules. evolving takagi sugeno (eTS) is proposed in \cite{eTS} with the concept of rule potential being an evolving version of mountain clustering method. \cite{eTS} is extended for classification problem in \cite{eClass} and termed evolving classifier (eClass). EFS has grown rapidly where a high number of works have been proposed in the literature. Recent survey of this area can be found in \cite{SkrjancIglesiasLughoferGomide19}. 

The area of EFS has been extended from a single model to an ensemble model in \cite{eENSEMBLE} using eTS as a base classifier. Nevertheless, this work is designed under a static ensemble structure which does not adapt to concept drifts of data streams. Parsimonious Ensemble (pENsemble) is proposed in \cite{PratamaPedryczLughofer18} to address data stream classification problems. pENsemble features a fully flexible structure where the ensemble structure generation is automated using the drift detection method and the ensemble pruning mechanism while a base learner makes use of parsimonious classifier (pClass) having a self-evolving network structure \cite{pClass}. pENsemble+ is proposed in \cite{pENsemble+} where it extends pENsemble with an online active learning method and an ensemble merging mechanism for online tool condition monitoring problem. Recently, another variant of evolving ensemble fuzzy neural network is proposed in \cite{OBEFS} putting forward the idea of online bagging for ensemble construction. The concept of ensemble classifier can be implemented using the stacked generalization principle making possible to configure an ensemble classifier in a deep structure. Such EFS work is realized in \cite{DSSCN} where it utilizes evolving stochastic configuration network (eSCN) as a base learner while the network depth is adjustable using a drift detection technique. This work is generalized in \cite{DEVFNN} using the feature augmentation technique for construction of deep neuro fuzzy structure. 

The data stream problem has also attracted research attention from the deep learning community. In \cite{AshfahaniP19}, the idea of autonomous deep learning (ADL) is proposed where it offers a self-evolving deep neural network under a different-depth network structure. In \cite{PratamaZAO019}, similar concept is proposed but it is developed from the framework of multi-layer perceptron network rather than the different-depth network structure. Recurrent Neural Network (RNN) is put forward in \cite{MUSE_RNN} to cope with the problem of data streams. The self-evolving concept is introduced in a single-layer denoising autoencoder (DAE) \cite{DEVDAN} extending the incremental feature learning concept in \cite{online_incremental_feature}. The concept of hedge back-propagation is proposed in \cite{ODL} to address the data stream problem using the different-depth network structure. In \cite{MLP}, the internet traffic classification problem is solved using the multi-layer perceptron network.

Although various solutions have been proposed to tackle the data stream problem, the vast majority of these works are not designed to cope with the large-scale data stream problem calling for distributed computing strategy to expedite the execution time. Note that the large-scale data stream problem differs from the conventional big data problem because it involves continuous information flow. A scalable PANFIS algorithm is proposed in \cite{za2019evolving} in which it implements PANFIS in \cite{PANFIS} under the distributed computing of Apache Spark and develops a model fusion technique. Nevertheless, Scalable PANFIS has not addressed the problem of continual data streams where data batches stream continuously. It is also not yet evaluated in the prequential test-then-train protocol as a standard evaluation procedure of data stream algorithms \cite{Gama10}. In \cite{ZainAPLP20}, a scalable teacher forcing network (ScatterNet) is proposed to overcome the bottlenecks of scalable PANFIS. It proposes a flexible ensemble classifier built upon a teacher-forcing fuzzy classifier and implemented under the distributed computing platform of Apache spark. Nevertheless, ScatterNet is a fully supervised algorithm imposing expensive labelling cost in the large-scale data stream environments. This issue requires an innovative algorithm handling the large-scale data stream problem under semi-supervised setting.

\section{Problem Formulation}
The large scale data stream problem is formalized as $B_1,B_2,...,B_K$ where $K$ denotes the number of data batches unknown in practice while $B_k$ labels the $k^{th}$ data stream with the size of $T$. Because of the speed of data generation, $T$ and $K$ are very large such that it is infeasible for a data stream to be processed under a traditional computing platform. $B_k$ comprises pairs of data points $\{x_t,y_t\}_{t=1}^{T}$ where $x_t\in\Re^u$ is an input data vector and $y_t\in\{l_1,l_2,...,l_m\}$ is a target class vector formed as a one-hot vector. Our model is simulated in the prequential test-then-train fashion \cite{BifetHolmesKirkbyPfahringer11} where it is forced to predict an unlabelled data batch $X_k\in\Re^{T\times u}$ while the predictive accuracy is measured independently per data batch. Once the target label $Y_k$ is available, model update and evolution is executed. 

The semi-supervised learning problem should be addressed in a large-scale data stream process due to the high manual labelling effort. That is, not all input samples $x_t$ can be paired by a target label $y_t$ meaning that the number of labels $T'$ in a data batch is much smaller than that of the batch size $T'<<T$. This issue requires particular strategies to retain the predictive accuracy in the case of scarcely labelled samples, while still being scalable to cope with the problem size. Another issue is observed in the aspect of structural complexity of the evolving models (base learners) which can become uncontrollably high due to the continual environments of data streams. The model fusion phase is needed after processing each data stream $B_k$ without compromising the predictive quality.      

The Spark environment consists of two parts: 1. the Spark's core; 2. the programming interface core. The Spark core executes the instructions of the programming interface core using its low level library. A large-scale data stream $B_k$ is processed in three steps. The first step is to store it in the memory cluster in the form of Spark's data frame. The second step is to divide it into $P$ data partitions and to distribute these to a number of computing nodes $N_o$. The last step is the consolidation phase where the distributed results are aggregated into a single result.

\begin{figure}[htbp]
	\begin{centering}
		\includegraphics[width=11cm]{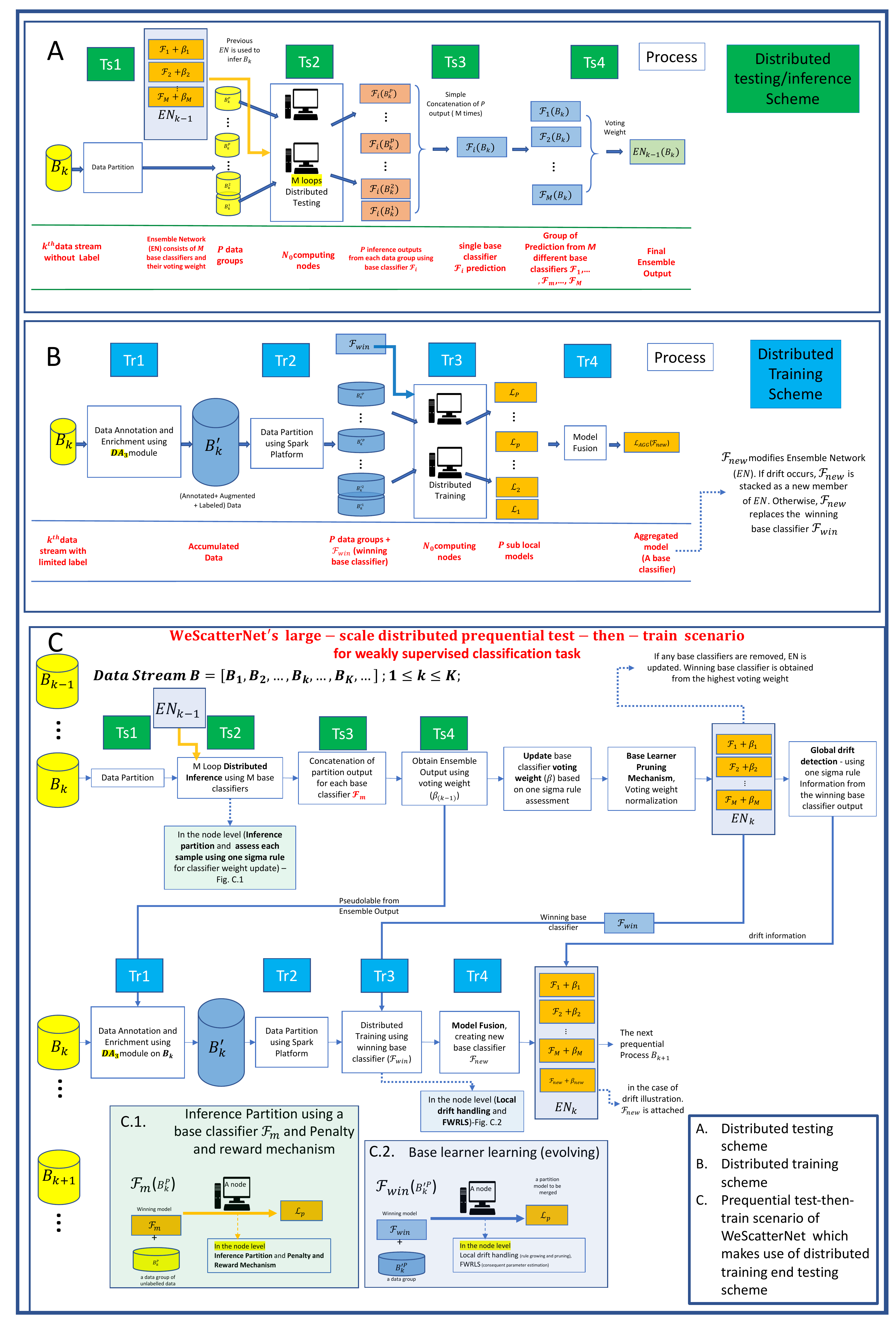}
		\par\end{centering}
	\caption{WeScatterNet's learning policy: WeScatterNet adopts a fully distributed training and testing process where they are executed in the distributed computing strategy of Apache spark. The distributed computing phase runs over $M$ base learners leading to an individual base learner prediction over a data batch $B_k$. Each base learner output is combined to produce the final ensemble output. The distributed training strategy occurs when adjusting the voting weight of a base learner (Step C.1) and fine-tuning the winning model (Step C.2) while $DA^3$ mechanism, the drift detection strategy and the model fusion step are carried out in the single node. The model fusion strategy combines $P$ models delivered from the distributed computing strategy into a compact model. The model is inserted as a new model if a drift is detected or replaces the winning model if a stable phase is returned.}
	\label{WeScatterLP}
\end{figure}

The large data stream $B_k$ is partitioned into $P$ data groups. A $B_k^p$ data group is processed by the worker node while its distribution is controlled by the driver node. It induces $P$ distributed models having to be aggregated into a single model to avoid the explosion of model's parameters. The consolidation phase plays a key role in the continual environments where a continuous arrival of data streams $B_k$ is expected. Furthermore, the consolidation phase should not compromise the overall model accuracy. The learning framework of WeScatterNet is pictorially illustrated in Fig. \ref{WeScatterLP} consisting of three sub figures. Fig. 2(A) shows the distributed testing procedure of WeScatterNet under Apache spark, Fig. 2(B) portrays the distributed training procedure of WeScatterNet under Apache spark and Fig. 2(C) depicts the distributed training and testing procedures of WeScatterNet.  

\section{Learning Procedure}
The learning procedure is described in Algorithm \ref{Pseudocode:LearningPolicyWeScatterNet} and is pictorially shown in Fig. \ref{WeScatterLP} (A),(B),(C).
WeScatterNet is capable of initiating its learning process from scratch without any predefined network structure. Its structural learning process encompasses the rule growing and pruning mechanisms based on the distributional variation of data streams \cite{ZainAPLP20}. It is simulated under the conventional prequential test-then-train protocol \cite{BifetHolmesKirkbyPfahringer11} where a model is forced to predict the incoming data batch $B_k$  before utilizing it for model updates. It is worth noting that only a small fraction of labelled data samples (see Tr1 process in algorithm \ref{Pseudocode:LearningPolicyWeScatterNet}) are made available for model updates.

\begin{algorithm}[!htbp]
	\caption{Learning Policy of WeScatterNet}\label{Pseudocode:LearningPolicyWeScatterNet}

	\begin{algorithmic}
		\STATE \textbf{Input} : Data Stream $B=[B_1,B_2,...,B_K,...]$
		\STATE \textbf{Ouput} : Evolution of Ensemble Network ($EN_1,EN_2,...,EN_K,...$) 
		\\  \textit{Initialization Process}:
		\\ $\pazocal{F}_1$ : Generate first model using $B_1$;
		$EN_{1} \leftarrow \pazocal{F}_1$
		\\  \textit{Loop Process}:
	
		\FOR {$k = 2$ to $K$}
	
		\STATE \textbf{I - Distributed Testing Fig. \ref{WeScatterLP} part. A} 
		\\ \textbf{Ts1:} Partition $B_k$ into $P$ partition; \textbf{Ts2+Ts3:} Perform $M$ loops distributed inference using $EN_{k-1}$. For each loop, concatenate all $P$ inference output $\pazocal{F}_i(B_k^p)$ as a base classifier's prediction ($\pazocal{F}_i(B_k)$) \textbf{Ts4:} Perform final ensemble inference ($\hat{Y}=EN_{k-1}(B_k)$)
		\\ \textbf{II- Update Base Classifier Weight ($\beta_k$) using Penalty and Reward Mechanism}
		\\ \textbf{III- Base Learner Pruning Mechanism}
		\\ If Condition eq.(\ref{clasprune}) is met, prune inconsequential base learners
		\\ \textbf{IV- Global Drift Detection}
		\\ \textbf{V - Distributed Training Fig. \ref{WeScatterLP} part.B}
		\\\textbf{Tr1:} Form Data Annotation and Enrichment, $B_k^{'}=DA_3(B_k^{labelled},B_k^{augmented},\hat{Y})$; \textbf{Tr2:}  Data Partition of $B_k^{'}$ using Spark;  \textbf{Tr3:}  Distributed training using $\pazocal{F}_{win}$; \textbf{Tr4:} Model Fusion forming new base classifier $\pazocal{F}_{new}$
		\\ \textbf{VI - Update $EN_k$}
		\\If drift is detected in \textbf{IV}, $\pazocal{F}_{new}$ is added to $EN_k$. Otherwise $\pazocal{F}_{new}$ replaces $\pazocal{F}_{win}$

	   \ENDFOR

	\end{algorithmic}

\end{algorithm}

WeScatterNet's learning protocol starts from the $DA^3$ module performing the self-labelling step of unlabelled samples and enriching the labelled samples. The penalty-and-reward step is carried out based on the compatibility of a model to the incoming sample. A reward is granted if a stream sample is sufficiently covered by the model from the sigma rule perspective, whereas a penalty is imposed if it is not sufficiently covered. In this sense, models representing an older, currently non-adequate state are down-weighed in the aggregation process for yielding the final model output. A drift detection method is executed to determine the need for an addition of a new model (for representing the drifted state). It is crafted from the Hoeffding's bound examining the statistics of the covariate. A drift leads to the insertion of a new model whereas the winning model (that one closest to the current sample) is updated when no drift is signalled. WeScatterNet presents an extension of ScatterNet \cite{ZainAPLP20} handling the semi-supervised learning problem within large-scale data stream environments.    

\subsection{Inference Procedure of WeScatterNet}
WeScatterNet makes use of the teacher forcing concept in which the hyper-plane membership function is exploited. It differs from the common TS fuzzy system \cite{TakagiSugeno85} where the rule premise is parameterized. It enables the rule consequent expressed in the form of a weighted linear combination of input attributes to form the rule premise, thereby leading to substantial reduction of network parameters. 
The membership degree of a data sample to the $i-th$ rule is defined in terms of the point-to-hyperplane distance \cite{PALM} as follows:
\begin{equation}\label{dist}
    d_{i,m}^{o}=\frac{|y_t^o/\hat{y}_{t-1}^o-x_e W_{i,m}^{o}|}{\sqrt{1+\sum_{j=1}^{u+1}W_{i,m,j}^o}}
\end{equation}
where eq.\eqref{dist} labels a distance between a data point and the $i-th$ rule of the $m-th$ base learner. $\hat{y}_{t-1}^o,y_{t}^{o}$ of eq.\eqref{dist} respectively denote the previously predicted output and the desired output at the $t^{th}$ (the current) time instant. $W_{i,m}^o\in\Re^{(u+1)}$ stands for the $o^{th}$ hyperplane of $i^{th}$ rule of $m^{th}$ base learner while $x_e=[1,x]$ is an extended input vector. Note that the symbol $/$ describes $y_t^o$ or $\hat{y}_{t-1}^o$ is applied alternately. The essence of the teacher forcing concept lies in the fact that the target variable $y_t^o$ is injected in the training process but replaced with $\hat{y}_{t-1}^o$ in the testing phase. The training process mixes both the desired and previously predicted output with the equal proportion. The use of $\hat{y}_{t-1}^o$ 
functions as some sort of an internal memory guiding the current prediction. It is perceived as a recurrent link as is also achieved in recurrent neural networks \cite{Mandic01,PALM}.  The firing strength of a rule is formalized by using the concept of the hyper-plane membership function:
\begin{equation}\label{firstrength}
    h_{i,m}^{o}=exp(-\frac{\gamma d_{i,m}^{o}}{\max_{i=1,...,R_{m}}d_{i,m}^{o}})
\end{equation}
where eq.\eqref{firstrength} denotes the firing strength of the $i-th$ rule. $R_m$ is the number of rules of the $m-th$ base learner. As with the conventional TS fuzzy system, the defuzzification process is implemented using the weighted average operation. 
\begin{equation}\label{output}
\hat{y}_o=\frac{\sum_{i=1}^{R_{m}} h_{i,m}^{o} x_e W_{i,m}^o}{\sum_{i=1}^{R_{m}} h_{i,m}^{o}}    
\end{equation}
where eq.\eqref{output} denotes the local output of a base learner. The predicted class label of a base learner is resulted from the maximum output of $O$ classes $\hat{o}=\max_{o=1,...,O}\hat{y}_o$. It can be seen as the MIMO architecture where every class possesses its own local sub-model, according to the one-versus-rest classification scheme \cite{HastieTibshiraniFriedman09} (leading to an indicator based regression task on $\{0,1\}$ (per class) for the consequent parameters, estimable in incremental manner with RLS and spin-offs, see below). Each base learner is assigned a specific voting weight $\beta_m$ (see subsequent section for assignment), which influence the final predicted output as being drawn from a weighted majority voting scheme across the $M$ base learners. 

\subsection{Penalty and Reward Mechanism of WeScatterNet}
WeScatterNet is constructed from the ensemble concept in which each base learner outputs its own local prediction. The weighted voting scheme is applied to infer the final prediction of WeScatterNet in which the voting weight is adjusted using the penalty-and-reward technique. A reward is granted if a base learner provides sufficient coverage of a data point and vice versa. That is, the compatibility test is performed with respect to the one-sigma rule as follows:
\begin{equation}\label{compatibility}
    \max_{i=1,...,R_m}h_i^{m}\geq 0.6065
\end{equation}
where eq.\eqref{compatibility} exhibits a penalty and reward condition. $R_m$ denotes the number of fuzzy rules of the $m^{th}$ base learner while $h_i^{m}$ stands for the firing strength of the $i^{th}$ rule in the $m^{th}$ base learner. Note that $0.6065$ comes from the normal distribution assumption notably the one-sigma rule meaning that a sample lies in very close proximity of the rule and a threshold selected from any unimodal distribution covers the majority of data points. The one sigma rule is chosen rather than the two sigma rule to avoid too lenient condition leading to excessive rewards and to assure sufficient coverage of data samples. 
The $m$-th base learner is assigned with a voting weight $\beta_m$ where a reward augments the voting weight:
\begin{equation}\label{reward}
    \beta_m=\min{(\beta_m*(1+fac),1)}
\end{equation}
where eq.\eqref{reward} is a reward mechanism augmenting the voting weight of a base learner. $fac\in[0,1]$ is a predefined constant. Eq.(\ref{reward}) is triggered if eq.(\ref{compatibility}) is satisfied. On the other hand, a penalty is carried out by diminishing the voting weight $\beta_m$, if eq.(\ref{compatibility}) is violated as follows:
\begin{equation}
\label{decrease_bm}
    \beta_m=\beta_m*fac
\end{equation}
where eq.\eqref{decrease_bm} denotes a penalty operation diminishing the voting weight of a base learner. The penalty and reward mechanism is to mirror the relevance of a base classifier to the current concept. An outdated base learner should have a low voting weight, thereby being ignored during the final decision making process. This is granted through the usage of eq.\eqref{decrease_bm}, as it decreases the influence weight of the base learner $\beta_m$ in the final output prediction whenever eq.(\ref{compatibility}) is violated, i.e. the base learner does not cover the sample well, hence the latter reflects a kind of 'drifted situation', which is not included as sub-model (rule) in the base learner. Thus, this base learner is expected to deliver an inaccurate prediction due to risk of extrapolation. 
The compatibility test is applied here rather than the accuracy vector \cite{AshfahaniP19} due to the scarcity of labelled samples, which makes an accumulated accuracy based on ahead-predictions of the base learner insignificant, thus not well representative. This module is executed in the distributed fashion across $P$ data partitions. The voting weight of the m-th base learner is aggregated from $P$ sub-models induced by $P$ data partitions of Apache Spark $\beta_m=\frac{\sum_{p=1}^{P}\beta_m^p}{P}$.
Then the overall classification output over all base learners is calculated as follows
\begin{equation}\label{global_output}
    \hat{Y}_o=\sum_{m=1}^{M}\beta_m\hat{Y}_m^o;\quad \hat{C}=\max_{o=1,...,O}\hat{Y}_o
\end{equation}
where $\hat{C}$ of eq.\eqref{global_output} stands for the final predictive output of WeScatterNet while $\hat{Y}_m$ of eq.\eqref{global_output} denotes the predictive output of the $m^{th}$ base classifier. 

\subsection{Global Drift Handling Mechanism}
WeScatterNet implements the global drift detection mechanism determining the learning actions to be undertaken. A new base classifier is amalgamated if a drifting distribution is identified while the stable case induces the adjustment of a winning base classifier having the highest voting weight. The cutting point $cut$ is solicited and reveals the switching point of data distributions $\hat{X}+\epsilon_X\leq\hat{A}+\epsilon_A, A\in\Re^{cut}$. $\hat{A},\hat{X}$ stand for the statistic of data partitions $A,X$ respectively. Only three candidates of the cutting point are considered here, e.g., $25\%,50\%,75\%$ to avoid false alarms. $\epsilon_{A,X}$ is the Hoeffding's bound formalized in eq.\eqref{Hoeffding's bound}.
\begin{equation}\label{Hoeffding's bound}
    \epsilon_{A,X}=(b-a)\sqrt{\frac{size}{2*cut*T}ln(\frac{1}{\delta})}
\end{equation}
where $size$ stands for the size of data partition of interest and $\delta$ denotes the significance level. Note that the significance level is inversely proportional to the confidence level $1-\delta$. $a,b$ denote the minimum and maximum points of input data samples $X_k$ seen so far. The cutting point $cut$ signifies the increase of population mean and leads to the concept drift. Two data partitions, namely $A\in\Re^{cut},C\in\Re^{T-cut}$, are formed. A drift condition is signalled if the null hypothesis, given by $|A-C|<\epsilon_{A,C}$, is rejected (as then a significant difference in the two partitions is observed). The concept drift condition calls for a new base learner to cover a new concept. A new base classifier is created from scratch and initialized with the winning base classifier. That is, the distributed computing strategy to induce a new base classifier refines the winning base classifier to be more fine-grained to describe the drift situation than it was before the model's update. The use of the winning base classifier as an initial classifier is reasonable because it has the closest relationship to the new concept.  
On the other hand, the stable concept occurs if the null hypothesis is not rejected. Such a case requires the winning base classifier (that one which is closest to the current sample) to be fine-tuned (following the conventional sequential, incremental learning concept) without adding a new base classifier to an ensemble structure. 

\subsection{Base Learner Pruning Approach}
The base learner pruning mechanism is incorporated in the WeScatterNet to alleviate the structural complexity. This mechanism checks the voting weight of a base learner where an inconsequential base learner having low voting weight is removed. Note that the voting weight is adjusted via the penalty and reward scheme where a low voting weight indicates low relevance to the current concept (as sample is not well covered) thus the learner should play little role to the final classification decision. The condition for removal of the $i^{th}$ base learner is formalized as follows:
\begin{equation}\label{clasprune}
    \beta_i\leq \mu_\beta-\sigma_\beta
\end{equation}
where eq.\eqref{clasprune} pinpoints the base learner pruning condition. $\mu_\beta,\sigma_\beta$ respectively denote the mean and standard deviation of the voting weights across all $M$ base learners. This approach follows the statistical process control concept as discussed in \cite{Gama10}, looking for typically low weights among all learners. It is worth mentioning that the base classifier weight is adjusted using the dynamic penalty and reward mechanism meaning that learner's importance is adjusted dynamically in respect to the recent context. In other words, the voting weight of a base learner reflects the importance of a base learner. Eq.(\ref{clasprune}) reveals an inconsequential base classifier which plays an insignificant role during its lifespan. Such base classier should be discarded to suppress the complexity of the ensemble classifier to a low level. 

\subsection{Base Learner Learning Approach}
After the condition of the current data stream is examined, either the addition of a new base learner or the adjustment of the winning base learner is performed in a distributed fashion --- Both call for the base learner update procedure (including rule evolution and pruning), which autonomously addresses drifts in a stream and is described in the following. 
\subsubsection{Local Drift Handling via Rule Growing and Pruning}
The base learner of WeScatterNet adopts the self-evolving paradigm where its rules are automatically generated and pruned with respect to variations of the data distribution. It utilizes an extension of the network significance method \cite{AshfahaniP19} derived from the bias variance decomposition. A new rule is added in the case of a high bias (under-fitting) whereas an outdated node is removed in the case of a high variance (over-fitting). The network bias and variance are crafted from $Bias=(y-E[\hat{y}])^2$ and $Var=(E[\hat{y}^2]-E[\hat{y}]^2)$.    

The expected output $E[\hat{y}]$ is defined as $\sum_{i=1}^{R} W_i\int_{-\infty}^{\infty}h_i(x;W_i)p(x)dx$. Solving the integral of hyperplane membership function is difficult. We assume the maximum membership degree of $h_i(x;W_i)=1$. Under the normal distribution assumption, the expression of the expected output is established in eq.\eqref{expectedoutput}.
\begin{equation}\label{expectedoutput}
    E[\hat{y}]=\sum_{i=1}^{R}\mu_e W_i
\end{equation}
where $\mu_e=[1,\mu]\in\Re^{u+1}$ denotes the expanded mean over all input features. Eq.(\ref{expectedoutput}) enables the establishment of network bias and variance. The term $E[\hat{y}^2]$ is solved by assuming the i.i.d condition navigating to $E[\hat{y}^2]=E[\hat{y}]*E[\hat{y}]$. The evaluation of network bias and variance from the training and validation errors is impossible here because of scarcely labelled samples. Furthermore, this approach complicates the evaluation of network statistical contribution in the streaming context because it depends on the testing error obtained in the next data batch. The probability density function $p(x)$ can be estimated by Gaussian Mixture Model (GMM) \cite{PratamaAH19} to cope with complex data distribution. This approach is, however, computationally expensive while often being unstable to handle the high input dimension. The forgetting factor here is integrated instead to keep pace with rapidly changing data distributions $p(x)_k\neq p(x)_{k+1}$. It is integrated in the recursive mean calculation as follows:
\begin{equation}\label{weightedmean_withforgetting}
    \mu_t=\mu_{t-1}+(f_t/F_t)(X_t-\mu_{t-1})
\end{equation}
where $F_t=F_{t-1}+f_t$. $F_t=t$,$f_t=1$ in the case of no forgetting while eq.\eqref{weightedmean_withforgetting} reflects the mean of normal distribution. The forgetting factor $f_t$ is calculated based on the drift rate of the data stream as $f_t=\exp{(-Rate_k)}$ and linearly scaled to the range of $f_t\in[0.9,1]$ to allow a smooth forgetting. The maximum forgetting occurs in the case of $0.9$ while no forgetting is applied at all in the case of 1. 
The point of interest in eq.(\ref{weightedmean_withforgetting}) is that drift handling is addressed in the concept mean while leaving aside the drift in the concept variance. It aims to keep pace with the core of changing data distributions rather than to enlarge its zone of influence which may result in lack of specificity. In addition, the expression of the expected output in eq.\eqref{expectedoutput} is not a factor of concept variance which can be safely ignored here. $Rate_k$ stands for the drift rate following the definition of the drift rate in \cite{charconceptdrift} where it is expressed in eq.\eqref{drift_rate}.
\begin{equation}\label{drift_rate}
    Rate_k=\lim_{\Delta\rightarrow\infty}\Delta D(k-0.5/\Delta,k+0.5/\Delta)
\end{equation}
where $\Delta D(.)$ stands for the total variation distance between two distributions. 
It is worth mentioning that the two distributions result from the two non-overlapping data groups. That is, $B_k^p$ is halved to form two data groups. The use of forgetting factor here is to meet the so-called sweet path \cite{charconceptdrift} where the high-bias-low-variance model is generated in the case of high drift rate whereas the low-bias-high-variance model is induced in the case of low drift rate.  

The paradigm of statistical process control \cite{Gama10} is adopted in the rule growing and pruning phases. The key difference lies in the use of bias and variance directly rather than the conversion of binomial distribution while having flexible confidence factor. 
\begin{equation}\label{growing}
    \mu_{t}^{bias}+\sigma_{t}^{bias}\geq \mu_{min}^{bias}+k_1\sigma_{min}^{bias}\rightarrow Growing
\end{equation}
\begin{equation} \label{pruning}
    \mu_t^{var}+\sigma_t^{var}\geq\mu_{min}^{var}+2*k_2\sigma_{min}^{var}\rightarrow Pruning
\end{equation}
where eq.\eqref{growing} and eq.\eqref{pruning} indicate the rule growing and pruning conditions. $\mu_t^{bias},\mu_t^{var},\sigma_t^{bias},\sigma_t^{var}$ denote the average of the bias, the average of the variance, the standard deviation of the bias, the standard deviation of the variance up to the current $t-th$ observation respectively, which can be recursively calculated. 
On the other hand, $\mu_{min}^{bias},\sigma_{min}^{bias},\mu_{min}^{var},\sigma_{min}^{var}$ denote the minimal values of the average of the bias and variance error and of the standard deviation of the bias and the variance error seen so far (over each update cycle). These are reset if eq.(\ref{growing}) or eq.(\ref{pruning}) is observed. The factor $2$ is inserted in eq.(\ref{pruning}) to prevent direct pruning after adding since the network variance naturally increases with the addition of new nodes but gradually reduces as the next observations come across. Note that $\mu_t^{bias},\mu_t^{var}$ have nothing to do with \eqref{weightedmean_withforgetting}. 

The normal distribution assumption is sightly relaxed here by applying adaptive confidence factors $k_1=1.25\exp{(-Bias^2)}+0.75$ and $k_2=1.25\exp{(-Var^2)}+0.75$ thus leading to $k_1=k_2=[1,2]$. That is, a new node is generated easily in the case of high bias whereas the rule growing condition becomes tight in the case of low bias. On the other hand, the rule pruning module is active if the variance is high and vice versa, 

if eq.(\ref{growing}) is met, a new rule is grown where its parameters are assigned:
\begin{equation}\label{newparams}
    {W_{R+1}} = k_3*\mathds{1}_{(u+1) , m}; \quad Sup_{R+1} = 1;\quad \Omega_{R+1}= \omega \mathds{I}_{(u+1),(u+1)}
\end{equation}
where $k_3$, $\omega$ are predefined constant while $W_{R+1}$, $Sup_{R+1}$, $\Omega_{R+1}$ of eq.\eqref{newparams} are the hyperplane, support and covariance matrix of the new rule. Note that $\omega$ is set as a positive large value to induce sufficient correction factor in the covariance matrix. If eq.(\ref{pruning}) holds, the weakest rule is pruned:
\begin{equation}\label{rulepruning}
\min_{i=1,...,R}E[\hat{y}]=\min_{i=1,...,R}\mu_{e}W
\end{equation}
where eq.\eqref{rulepruning} signifies the rule pruning mechanism. The weakest node is defined as that having the smallest statistical contribution over past samples. In such a case, the support of the inactive rule is transferred to the winning rule $Sup_{win}=Sup_{win}+Sup_{pruned}$. If eq.(\ref{growing}) is violated, the support of the winning rule is incremented $Sup_{win}=Sup_{win}+1$, and its consequent parameters are updated by the modified FWGRLS method (also to handle noisy pseudo labels properly) as explained below. 

\subsubsection{Data Augmentation, Annotation and Auto-Correction ($DA^3$)}
WeScatterNet is developed to handle the semi-supervised learning scenario of large-scale data streams via $DA^3$ approach. $DA^3$ approach governs the parameter learning phase using the label augmentation, the automatic labelling of unlabelled samples and regularization. The overall  cost function consists of three components as follows:
\begin{equation}\label{loss}
    L_{overall}=L(\hat{y},y)+L(\hat{y}_{ps},y_{ps})+L(\hat{y}_{aug},y_{aug})
\end{equation}
where eq.\eqref{loss} denotes the overall loss function. The first term denotes the loss of originally labelled samples $L(\hat{y},y)$ followed by the loss of pseudo labels $L(\hat{y}_{ps},y_{ps})$ and augmented labels $L(\hat{y}_{aug},y_{aug})$ respectively. The augmented label $y_{aug}$ is a variation of originally labelled samples using a controlled noise while leaving the labels unaffected. That is, a small Gaussian noise with zero mean $N(0,0.001)$ is injected to the originally labelled samples here thus producing the corrupted version of original samples. Since the augmented samples are generated from the originally labelled samples, they are deemed to be clean.  

The pseudo label is generated by the self-labelling mechanism of unlabelled samples elicited from the final predicted label of ensemble classifier. This step ought to be carefully carried out because of the risk of noisy pseudo-label. That is, a wrong label is fed thereby resulting in significant performance compromise. A pseudo-label, $y_{ps}$, is generated if all base classifiers are confident with their own predictions and have agreeable predictions meaning that all ensemble members produce the same class label. This concept is formalised as follows:
\begin{equation}\label{prediction}
    \min_{m=1,...,M} conf_{m} \geq 0.55 \quad \& \quad \hat{o}_1=\hat{o}_2=,...,=\hat{o}_M  
\end{equation}
where eq.\eqref{prediction} indicates the pseudo-label assignment condition. $conf_{m}=\frac{\hat{y}_1^m}{\hat{y}_1^m+\hat{y}_2^m}$ denotes the confidence level of the $m^{th}$ base classifier while $\hat{o}_m$ labels the predicted class label of $m^{th}$ base classifier. $\hat{y}_1^m, \hat{y}_2^m$ respectively stand for the highest predictive output and the second highest predictive output of the $m^{th}$ base classifier. $conf_m$ exhibits whether or not a base classifier is certain to its prediction. $conf_m\approx0.5$ implies confused prediction since its prediction is not conclusive to a particular class. In other words, a prediction lies nearby the decision boundary. By extension, eq.(\ref{prediction}) requires agreeable base classifiers. One can envisage each ensemble member describes local data space. This condition implies a data point sitting in the overlapping area of all ensemble members. $y_{ps}$ is assigned as that $\hat{o}_m$ if eq.(\ref{prediction}) is satisfied. The specific regularization method is in addition integrated to address the noisy pseudo-label problem where the main goal is to prevent important rules to move away from its ideal location as induced by the original label while allowing less important rule to embrace the pseudo-label. It is made possible with the use of local learning where each rule is adjusted separately. Any changes to one rule incurs low influence on the convergence of other rules.      

\subsubsection{Recursive Learning of Consequent Parameters with a modified version of FWGRLS}
The consequent parameters $W_i$ are recursively estimated via the fuzzily weighted generalized recursive least square (FWGRLS) method offering an extension of fuzzily weighted recursive least square (FWRLS) method with the use of a weight decay term to enhance the generalization power. The FWGRLS method has its root in the GRLS method \cite{pratama2013genefis} where the weight decay term is introduced in the cost function of the RLS method. The key difference between the FWGRLS method and the GRLS method exists in the local learning scheme where the FWGRLS method updates each rule separately with its own inverse Hessian matrix $\Omega_i=(X^TQ_iX)^{-1}$, with $Q_i$ the weighting matrix including the membership degrees of the samples to the $i$th rule.
These degrees makes the matrix 'local', as samples lying far from rule $i$ receive low weights and are thus hardly respected in the estimation of the consequents. The local learning scheme is important in a self-evolving context since any growing, deletion and update on one rule does not affect the consequent parameters of the other rules (no disturbance of optimality etc.) (see Chapter 2 in \cite{LughoferBook11} for a detailed analysis). 
The FWGRLS method is written as follows:
\begin{equation}
\label{recursive_weight_update}
    W_i^t=W_i^{t-1}-\alpha\Omega_i\nabla\phi(W_i^{t-1})+K_i^{t}(y^t-W_ix_e) 
\end{equation}
\begin{equation}\label{covariance_update}
    \Omega_i^{t}=\Omega_i^{t-1}-K_i^tx_e\Omega_i^{t-1}
\end{equation}
\begin{equation}\label{KalmanGain_update}
    K_i^{t}=\Omega_i^{t-1}x_e(1+x_e\Omega_i^{t-1}x_e^{T})^{-1}
\end{equation}
where eq.\eqref{recursive_weight_update}, eq.\eqref{covariance_update} and eq.\eqref{KalmanGain_update} signify the hyperplane update strategy of the FWGRLS method. $\phi(W_i^{t-1})$
denotes the weight decay term of the $i^{th}$ rule regularizing the model update while $\alpha$ labels the predefined constant controlling the intensity of regularization and is assigned an extremely small value ($\alpha\approx 3*10^{-7}$). 
$K_i^{t}$ is the Kalman gain of the $i^{th}$ rule. The weight decay term $\phi(W_i^{t-1})$ steers the magnitude of the weight vector $W_i^{t-1}$, trying to keep them in a small range. This typically improves the generalization power \cite{Mackay92} \cite{LeungWongXu08} while achieving implicit dimension reduction affecting the consequent hyper-planes. A quadratic weight decay function is used here since it is one of the most widely used weight decay terms.
\begin{equation}\label{quadratic_weight_decay_term}\phi(W_i^{t-1})=\frac{1}{2}(W_i^{t-1})^{2};\quad \nabla\phi(W_i^{t-1})=W_i\end{equation}
where eq.\eqref{quadratic_weight_decay_term} stands for the quadratic weight decay term and its gradient. The quadratic weight decay term is capable of reducing the magnitude of the hyperplane proportionally to its current values. 

The original FWGRLS method is only applied for originally labelled samples and augmented samples.
The FWGRLS is modified to handle the noisy pseudo label samples. That is, the weight update formula is expressed:
\begin{equation}\label{newupdate}
 W_i^t=W_i^{t-1}-\alpha\Omega_i\frac{\sum_{n=1}^t h_{n,i}}{t-t'}\nabla\phi(W_i^{t-1}-W_i*)+K_i^{t}(y^t-W^{t-1}x_e)
\end{equation}
where eq.\eqref{newupdate} denotes the hyperplane update strategy when dealing with pseudo label. $t'$ stands for the time index where a rule is added in the rule base. The term $\frac{\sum_{n=1}^t h_{n,i}}{t-t'}$ aims to take into account the importance of $i^{th}$ rule in the model update. 
The higher the importance of a rule, the higher the regularization intensity becomes, which is in accordance to the weighted sampling strategy as integrated in the original FWGRLS following the local learning spirit. In other words, an important rule is hindered to accept the noisy pseudo label. Furthermore, the weight decay term here is defined as follows:
\begin{equation}\label{regularizer}
    \phi(W_i^{t-1}-W_i*)=\frac{1}{2}(W_i^{t-1}-W_i*)^2;\nabla\phi(W_i^{t-1}-W_i*)=(W_i^{t-1}-W_i*)
\end{equation}
where eq.\eqref{regularizer} denotes the modified weight decay term and its gradient when handling the pseudo label. $W_i*$ stands for the weight vector before receiving pseudo label. $(W_i^{t-1}-W_i*)$ functions as auto-correction mechanism in the case of noisy pseudo labels. That is, it forces the weight vector to be as close as possible to its prior values before receiving pseudo labels. It is seen from the negative sign of the update formula in \eqref{newupdate}. 
In other words, the regularization strategy freezes the parameters of important rules. This strategy is inspired by that in \cite{kirkpatrick2016overcoming} where it is extended to the fuzzy rule level rather than the synaptic level. Moreover, this strategy is incorporated in the realm of FWGRLS method rather than the back propagation method. It is worth noting that FWGRLS is an exact approximation of Least Square (LS) solution assuring convergence in a single update, because LS is a convex parabola function and eq.\eqref{recursive_weight_update} denotes a Gauss-Newton step, which converges in a single iteration for such parabolas. Eq.(\ref{newupdate}) and eq.(\ref{regularizer}) are applied only when seeing the pseudo label.

\subsection{Data Free Model Fusion}
The model fusion step is designed to keep the structural complexity at a reasonable level which might go to untenable level in the continual environments. It is a factor of the number of data streams $K$ which might be unbounded in practise. Specifically, the number of rules can go to the level of $P*K*R$ where $P,K,R$ respectively denote the number of data partitions, streams and rules in the distributed node. Furthermore, there exists the issue of accuracy where the model fusion step must produce at least the comparable level of accuracy as that in the single node environment. The local learning property of WeScatterNet provides flexibility in the distributed environment where every rule possesses its own inverse Hessian matrix and consequent vector (separately updated by the modified FWGRLS algorithm) which does not incur substantial accuracy degradation during the model fusion phase --- an issue which becomes apparent in the case of conventional global learning (most commonly used in literature). 

\subsubsection{Similarity Analysis}
The model fusion strategy is derived from the hyper-plane merging concept where similar hyper-planes can be merged into one without suffering from any accuracy drop. The similarity is calculated based on the distance $Dist$ and the dihedral angle $\Phi$ spanned by the two hyper-planes as follows:
\begin{equation}\label{dist2}
    Sim_1(i,i') = Dist(i,i')=(||W_i-W_{i'}||)/(||W_i+W_{i'}||)
\end{equation}
\begin{equation}\label{ang}
    Sim_2(i,i')=\Phi/\pi; \quad \Phi=arccos(\frac{a^Tb}{|a||b|})
\end{equation}
where eq.(\ref{dist2}) and eq.(\ref{ang}) respectively stand for the normalized distance of two hyperplanes and the dehideral angle between the two normal vectors $a=[W_{i,1},W_{i,2},...,W_{i,u},-1]$ and $b=[-W_{i',1},-W_{i',2},...,-W_{i',u},1]$. The vector $a$ is represented by the coefficient of the hyper-planes with respect to the input variables and $-1*y$ because of $W_{i,1}x_1+W_{i,2}x_2+...+W_{i,u}x_u-y=-W_{i,0}$. It reflects the normal vector of a hyperplane with $W_{i,0}$ as the intercept. Another vector $b$ is set to the opposite direction in order to correctly obtain the dehideral angle. A rule is merged to another rule if the following condition is met.
\begin{equation}\label{simcondition}
    Sim_1\leq k_4 \quad \& \quad Sim_2 \geq k_5 
\end{equation}
where eq.\eqref{simcondition} denotes the rule merging condition. $k_4$ and $k_5$ denote the predefined similarity thresholds and govern the intensity of merging process. The smaller the values of $k_4$ the less frequent the merging process is carried out and vice versa. On the other side, the higher the values of $k_5$ the less frequent the merging process is undertaken. 

\subsubsection{Merging Process}
The merging process is carried out by first removing inconsequential rules deteriorating the model's generalization if it is merged with other rules. Inconsequential rules are those possessing minor supports thus representing low variance direction of data distribution. The minimum support is capped at $2\%$ of the batch size $T$. 

The merging process is implemented in the $Z-best-rules$ fashion. That is, the best $Z$ rules are extracted with respect to the training accuracy in which other rules are merged to these $Z$ nodes by examining their angle and distance \eqref{simcondition}. That is, other rules are coalesced to one of those $Z$ rules having the highest similarity. In other words, the merging process is carried out in the greedy fashion rather than the one-to-one fashion. An online model selection is carried out to determine $Z$ here. 

\subsubsection{Online Model Selection}
The online model selection strategy aims to select the best $Z$ rules attaining tradeoff between accuracy and simplicity. $Z$ is not left as a hyper-parameter here leading WeScatterNet to be ad-hoc. Note that $Z$ plays vital role since it determines the number of rules in the rule base where other rules are blended into the best $Z$ rules. The candidate of $Z$ is selected as $3,5,8,10$. The underlying objective of the model selection is to obtain $Z$ minimizing both network bias and variance simultaneously while attaining the best classification performance.
\begin{equation}\label{OMS}
    \min_{Z=3,5,8,10}\frac{|Bias_{B_{sample}}*Var_{B_{sample}}|}{Acc_{B_{sample}}}
\end{equation}
where eq.\eqref{OMS} labels the condition of model selection. The candidate minimizing eq.(\ref{OMS}) is selected. Furthermore, eq.(\ref{OMS}) is evaluated using $B_{sample}$. That is, one data point per partition is drawn at random from labelled samples to form $B_{sample}$. Note that there are $P$ data partitions in total. \eqref{OMS} avoids the underfitting and overfitting conditions while attaining low empirical error at the same time. The model fusion approach is described in Algorithm \ref{Pseudocode:MergingProcedure}. 

\begin{algorithm}[!htbp]
	\caption{Model Fusion of WeScatterNet}\label{Pseudocode:MergingProcedure}

	\begin{algorithmic}
		\STATE \textbf{Input} : (1) Initial Model ($\pazocal{L}_{init}$ - $P$ base learners): generated from the winning base learner $\pazocal{F}_{win}$ using the training dataset $B_k^{'}$; (2) $B_{sample}$: Collection of instances, where each instance is taken randomly from labelled data partition $B_k^{p,labelled}$, ($B_{sample} \subset B_k$)
		\STATE \textbf{Ouput} : Aggregated model $\pazocal{F}_{new}$ 
		\\\textit{Initialization Process:}
		\\I \textbf{- Rules extraction}-extract all rules from $\pazocal{L}_{init}$. \\\qquad$\pazocal{L}_{extract}=(\pazocal{L}_1,...,\pazocal{L}_p,...,\pazocal{L}_P)$, where $\pazocal{L}_p=(Rules_1,...,Rules_R)$, forming 
		\\\qquad $O$ number of rules extracted from $\pazocal{L}_{init}$.
		\\II \textbf{- Assign all rules with the classification training performance}
		\\\qquad This value, $Classification_{training}$, is obtained from \\\qquad performance of each rule in $\pazocal{L}_p$ to the training dataset $B_k'$
		\\III \textbf{- Rules elimination} 
		\\\qquad\textbf{if} $Sup_{i}<0.02*\sum_{i=1}^OSup_{i}$ \textbf{then} $rules/nodes_i$ are deleted
		\\IV \textbf{- Sorting} $\pazocal{L}_{extract}$; $\pazocal{L}_{sort}=Sorting(\pazocal{L}_{extract},classification_{training})$  
		\\\qquad Obtain rules ranking by sorting $\pazocal{L}_{extract}$ using \\\qquad classification training performance $classification_{training}$
		\\V \textbf{- Online Model Selection} 
		\STATE \textit{Initialization: }
		\STATE $candidateZValue=3,5,8,10$;  
		\STATE $candidateModel$; $listPerformance=[ ]$; $count=0$
    	\FOR {$iVar$ in $candidateZValue$}
	    \STATE $count=count+1$
		\STATE \textbf{I - Obtain the dominant rules} ($\pazocal{L}_{dom}=\pazocal{L}_{sort}[1:iVar]$) 
		\STATE \textbf{II - Obtain the candidate rules} ($\pazocal{L}_{cand}=\pazocal{L}_{sort}[(iVar+1):O]$) 
		\STATE \textbf{III - Merging candidate rules into dominant rules} 
		\\\qquad $candidateModel[count]=\pazocal{L}_{merged}$ 
		\\\qquad\qquad (a) Compare each rule in $\pazocal{L}_{cand}$ and calculate $sim_1$ and $sim_2$ with \\\qquad\qquad all dominant rules in $\pazocal{L}_{dom}$
	    \\\qquad\qquad (b) If condition eq.(\ref{simcondition}) is met, merge candidate rule into \\\qquad \qquad the most minimum ($sim_1$)
        \STATE \textbf{IV - Calculate Score for each $\pazocal{L}_{merged}$} 
		\\\qquad$[Acc_{B_{sample}},Bias_{B_{sample}},Var_{B_{sample}}]=\pazocal{L}_{merged}(B_{sample})$
		\\\qquad$PerformanceValue=Bias_{B_{sample}}*Var_{B_{sample}}/Acc_{B_{sample}}$
		\\\qquad $listPerformance[count]=PerformanceValue$
	\ENDFOR
	\\ VI \textbf{- Select the best model for updating the Ensemble Network(EN)} 
		\\\qquad $Zindex_{selected}=\min listPerformance$
	    \\\qquad \textbf{Aggregated Model:} $\pazocal{F}_{new}=candidateModel[Zindex_{selected}]$

	\end{algorithmic}

\end{algorithm}

\section{Numerical Study}
The efficacy of WeScatterNet is numerically validated in two scenarios: large-size-small-number and small-size-large-number. The latter case refers to a large number of data batches while having small-size per batch. It aims to evaluate the advantage of model fusion scenario whether or not it remains scalable and retains high accuracy since it is triggered frequently in this case. The former one captures a case where a data stream is large in size whereas the number of stream is limited. It tests the scalability of the distributed training policy. Furthermore, the effect of label's proportion is studied here where the performance of WeScatterNet is evaluated under four label proportions, namely 10\%, 25\%, 50\%, 75\% as depicted in the Table \ref{tab:NumResultsWeScatterNetReg}. Ablation study is conducted to investigate the individual contribution of each learning module to the final numerical results. We also study the performance of WeScatterNet when implemented in the single node environment. This study is designed to evaluate the performance gap with the distributed implementation. Our numerical study is simulated under the prequential-test-then-train fashion where numerical results are calculated independently per data stream. 


\subsection{Dataset}
Six popular big data problems, namely Higgs \cite{Baldi}, Susy \cite{Baldi}, Hepmass \cite{Baldi}, RLCPS \cite{RLCPS}, KDDCup \cite{KDDCup} and PokerHand \cite{Dua2019}, are utilized to evaluate the performance of WeScatterNet where their properties are summed up in Table \ref{datasetproperties} Detailed characteristics of these problems are outlined as follows:

\begin{table}[htbp]
  \centering
  \caption{Dataset properties and their prequential setting}
  \scalebox{0.80}{
          \begin{tabular}{ccccccc}
        \toprule
        Dataset & \#IA    & \#C     & \#Instances  & Setting & Nbatch & NSampleBatch \\
        \midrule
        \multirow{2}[1]{*}{Higgs} & \multirow{2}[1]{*}{28} & \multirow{2}[1]{*}{2} & \multirow{2}[1]{*}{11,500K} & Large & 66    & 166.7K \\
              &       &       &       & Small & 198   & 55.6K \\
        \midrule
        \multirow{2}[0]{*}{Hepmass} & \multirow{2}[0]{*}{28} & \multirow{2}[0]{*}{2} & \multirow{2}[0]{*}{11,000K} & Large & 63    & 166.7K \\
              &       &       &       & Small & 189   & 55.6K \\
         \midrule
        \multirow{2}[0]{*}{Susy} & \multirow{2}[0]{*}{18} & \multirow{2}[0]{*}{2} & \multirow{2}[0]{*}{5,000K} & Large & 30    & 166.7K \\
              &       &       &       & Small & 90    & 55.6K \\
         \midrule
        \multirow{2}[0]{*}{RLCPS} & \multirow{2}[0]{*}{9} & \multirow{2}[0]{*}{2} & \multirow{2}[0]{*}{5,000K} & Large & 30    & 166.7K \\
              &       &       &       & Small & 90    & 55.6K \\
         \midrule
        \multirow{2}[0]{*}{KDDCup} & \multirow{2}[0]{*}{2} & \multirow{2}[0]{*}{2} & \multirow{2}[0]{*}{4,898K} & Large & 29    & 168.8K \\
              &       &       &       & Small & 87    & 56.29K \\
         \midrule
        \multirow{2}[1]{*}{Pokerhand} & \multirow{2}[1]{*}{10} & \multirow{2}[1]{*}{10} & \multirow{2}[1]{*}{1,025K} & Large & 6     & 170.8K \\
              &       &       &       & Small & 18    & 56.9K \\
        
         \midrule
         
        \bottomrule
        \end{tabular}%

  }

 	\label{datasetproperties}
\end{table}%

\noindent\textbf{Higgs}: it is an artificial classification problem whose goal is to classify a signal process leading to Higgs Boson particle. This problem is generated from the monte carlo simulation where the first 21 features describe the kinematic properties measured by the particle detectors in the accelerator. The last seven features are the high level features used to discriminate the Higgs Boson particles. 

\noindent\textbf{SUSY}: it is similar to the Higgs problem but the goal is to identify the signal process generating supersymmetric particles. It is generated by the Monte Carlo simulation where the first 8 features are the kinematic attributes while the last 10 features are the function of the first 8 features. 

\noindent\textbf{Hepmass}: the goal of this problem is to find an exotic particles via a large number of collisions. The point of interest is a particle with an unknown mass. It consists of 28 input attributes where 22 features are low-level features while the rests are high-level features. It is a binary classification problem to distinguish between the signal process from the background.

\noindent\textbf{PokerHand}: this problem is a multi-class classification problem with 10 classes where each record describes a hand comprising five playing cards drawn from a standard deck of 52. This problem has 10 input attributes where each attribute outlines a card to be either suit or rank. This problem is well-known for its non-stationary characteristic because of its dependencies on the card's orders.   

\noindent\textbf{KDDCup}: this problem describes intrusion detection problem whether or not an attack has occurred. It presents non-stationary properties because it simulates various types of network intrusion in the military environments and consists of 41 input attributes.

\noindent\textbf{RLCPS}: This problem is comparison problem of individual data collected in the course of several years from 2005 to 2008. It is a binary classification problem between 'match' or 'not match'. This problem presents 11 input attributes and over 5 million instances. Only around 20 K samples come from match category thereby leading to the skewed class distribution problem. 

\subsection{Baseline}
WeScatterNet is compared against two baselines, Scalable PANFIS \cite{za2019evolving} and ScatterNet \cite{ZainAPLP20} featuring the fully supervised learning approaches. Scalable PANFIS  is a generalized version of PANFIS \cite{PANFIS} for big data stream problem. It features an open structure under a single base classifier framework. ScatterNet is an ensemble classifier based on the teacher-forcing concept akin to WeScatterNet. It is, however, devised for the fully supervised learning environment. All algorithms are run under the Apache spark platform under the same computational resources. The hyper-parameters of WeScatterNet and baseline algorithms are selected as Table \ref{tab:TableOfParameters} and fixed for all simulations in this paper. All algorithms are executed in the same computational environments in order for execution time to be compared. Numerical results are produced by executing their published codes. The Source code of WeScatterNet can be completely downloaded from the following link : \href{https://github.com/ContinualAL/WeScatterNet.git}{\textit{\textbf{WeScatterNetCodeLink}}}.

\begin{table}[htbp]
  \centering
  \caption{Hyper-parameters of WeScaterNet and the Baseline algorithms}
  \scalebox{0.60}{
        \begin{tabular}{cllll}
    \toprule
    \multicolumn{1}{l}{\textbf{Algorithm}} & \textbf{parameters} & \textbf{Value} & \textbf{Module}  & \textbf{Descripition} \\
    \midrule
    \multicolumn{1}{c}{\multirow{8}[2]{*}{\begin{tabular}[c]{@{}c@{}}WeScatterNet \\/ScatterNet\end{tabular}}} & fac & 0.3/0.3   & Penalty and & follows eq (5) in the reward case,\\
    &  &    & Reward Mechanism   & and eq(6) in the penalty case.  \\
          & $\alpha_{drift}$ & 10\^(-3)/10\^(-3) & \multicolumn{1}{l}{Drift Detection} & controls the drift rate of the output  \\
          &  &    &  & in the global level  \\
          & $k_4$(dist)    & 0.4/0.4   & Merging & maximum distance two rules can be merged \\
          & $k_5$(angle)    & 0.6/0.6   & Merging & minimum angles two rules can be merged\\
          & $\gamma$ & 0.7/1.4   & local learning & control parameter which steers  \\
           &  &    &  &the degree of membership function  \\
          & $k_3$ & 0.2/0.4   & local learning & coefficient for weight initialization  \\
           &  &    &  & of the first sample  \\
    \midrule
    \multicolumn{1}{c}{\multirow{3}[2]{*}{ Scalable PANFIS }} & \multicolumn{1}{p{8.715em}}{kfs} & 0.05  & local learning & safety width \\
          & kgrow & 1     & local learning & growing node threshold \\
          & kprune & 0.25  & local learning & pruning node threshold \\
    \bottomrule
    \end{tabular}%
    }%
  \label{tab:TableOfParameters}%
\end{table}%

\subsection{Computational Environments}
The computational environments are configured under Nimbus Pawsey Supercomputing Centre Australia. The Spark cluster is constructed by using seven computing nodes consisting of one master node and six worker nodes. Each node has an identical specification as follows: 8VCPUs(cores), 32GB of RAM, and 40GB of hard-disk capacity, where Pawsey version of Ubuntu 20.04 Focal Fossa is installed. For software specification, we use Apache Spark version 3.0.0 and R version 3.6.3. Of 32 GB of RAM in each worker node, 24 GB of RAM is used for the Spark process, whereas the remaining RAM capacity is used for other background processes. Thus, for six worker nodes, the Spark cluster utilizes a total of 144 GB of RAM. For a driver node, we allocate only 8 GB of RAM. Since the traffic optimization is beyond the scope of our paper, we simply apply one executor per node for our distributed computing strategy. We start from one master node and one worker node to determine a suitable number of worker nodes and then increase the number of worker nodes until a point where reasonable execution time is attained, i.e., one master node and six worker nodes. We find that execution time decreases by increasing the number of worker nodes up to a point where reduction of execution time is no longer significant. 

\begin{table}[p]
  \begin{center}
  \caption{Numerical Results of Consolidated Algorithms under Large Setting and Small Setting: it is presented that our algorithm presents comparable result despite constrained by low number of labelled samples.}
  
     \scalebox{0.7}{
{\renewcommand{\arraystretch}{0.70} 
\begin{tabular}{llllll}  & &  &    &   &      \\ 
\toprule
\hline
Algorithm & Dataset   & \begin{tabular}[c]{@{}c@{}}Average \\accuracy per \\batch (\%)\end{tabular} & \begin{tabular}[c]{@{}c@{}}Avarage \\Training Time \\ per batch (s)\end{tabular} & \begin{tabular}[c]{@{}c@{}}Avarage \\Testing Time \\ per batch (s)\end{tabular} & \begin{tabular}[c]{@{}c@{}}Average \\Number of \\Model\end{tabular}  \\ 
\midrule
\multirow{6}{*}{\begin{tabular}[c]{@{}c@{}}WeScatterNet with \\Regularization\\\textbf{using 25 percent}\\of labeled data\\LARGE SETTING\end{tabular}}

& Higgs     & 63.60& 19.57 & 17.51 & 2.51  \\
& Hepmass   & \textbf{83.44}& 40.59 &7.44  & 1  \\
& Susy      & 75.67& 32.07  &11.27  &  2 \\
& RLCPS     & 99.64& 52.66 & 4.79 & 1\\
& KDDCup    & 99.53& 65.53 & 9.01 & 1  \\
& PokerHand & \textbf{50.13}& 8.59 & 8.33  & 1  \\

\hline
\multirow{6}{*}{\begin{tabular}[c]{@{}c@{}}ScatterNet \\LARGE SETTING\end{tabular}}  
& Higgs     & \textbf{63.75}& 8  & 16.57 & 2.48 \\
& Hepmass   & 83.43& 8.04   & 14.09 & 2.32 \\
& Susy      & 75.43& 6.7& 12.86 & 2.55 \\
& RLCPS     & 99.65& 5.56   & 4.23  & 2.01 \\
& KDDCup    &\textbf{99.59}& 10.04  & 8.13  & 1    \\
& PokerHand & \textbf{50.13}& 9.76   & 7.73  & 1    \\ 
\hline
\multirow{6}{*}{\begin{tabular}[c]{@{}c@{}}Scalable PANFIS \\LARGE SETTING\end{tabular}} 
& Higgs     & 63.29  & 76.19  & 5.33  &    1  \\
& Hepmass   & 83.26 & 74.20   & 5.67  &    1  \\
& Susy      & \textbf{75.68} &  36.77  & 3.88  &    1  \\
& RLCPS     & \textbf{99.78} & 15.53   & 2.92  &    1  \\
& KDDCup    & 99.46 & 147.44   & 7.03  &    1  \\
& PokerHand & 50.04 & 31.57   & 4.25  &     1 \\
\hline

\hline
\multirow{6}{*}{\begin{tabular}[c]{@{}c@{}}WeScatterNet with \\Regularization\\\textbf{using 25 percent}\\of labeled data\\SMALL SETTING\end{tabular}}    
& Higgs     &\textbf{63.26} &6.43&5.1&2.01\\
& Hepmass   & \textbf{83.45} &9.24 &2.748& 1 \\
& Susy      & \textbf{75.7} &7.01 &2.25&  1 \\
& RLCPS     &99.64& 9.19 &1.95 & 1 \\
& KDDCup    & 99.41 &13.6 &3.23&  1 \\
& PokerHand &\textbf{50.11}& 4.67 &3.07 &  1 \\
\hline

\hline
\multirow{6}{*}{\begin{tabular}[c]{@{}c@{}}ScatterNet \\SMALL SETTING\end{tabular}}  
& Higgs     & 63.09& 3.46   & 5.48  & 2.3  \\
& Hepmass   & 83.22& 3.51   & 3.7   & 1.57 \\
& Susy      & 74.95& 3  & 4.41  & 2.29 \\
& RLCPS     & 99.64& 2.63   & 1.82  & 1    \\
& KDDCup    & 99.28& 4.24   & 3.04  & 1    \\
& PokerHand & 50.09& 5.25   & 2.97  & 1    \\ 
\hline
\multirow{6}{*}{\begin{tabular}[c]{@{}c@{}}Scalable PANFIS \\SMALL SETTING\end{tabular}} 
& Higgs     &  61.97 & 25.37   & 2.09  &    1  \\
& Hepmass   & 82.86 & 24.52   & 2.21  &    1  \\
& Susy      & 75.51 & 12.89   & 1.64  &    1  \\
& RLCPS     & \textbf{99.70} & 5.65   &  1.21 &    1  \\
& KDDCup    & \textbf{99.46} & 147.44   & 7.03  &    1  \\
& PokerHand & 49.96 & 10.11   &  1.67 &     1 \\

\hline

\bottomrule
\end{tabular}
}
}
    \centering{}
  \end{center}

  \label{tab:NumResult}%
\end{table}

\begin{table}[p]
  \begin{center}
  \caption{Comparison with Single Node Algorithms: it is presented that our algorithm produces competitive results although it is constrained by lack of labelled samples and distributed computing strategy. It produces much faster execution time.}
  
     \scalebox{0.7}{
{\renewcommand{\arraystretch}{0.70} 
\begin{tabular}{llllll}  & &  &    &   &      \\ 
\toprule
\hline
Algorithm & Dataset   & \begin{tabular}[c]{@{}c@{}}Average \\accuracy per \\batch (\%)\end{tabular} & \begin{tabular}[c]{@{}c@{}}Avarage \\Training Time \\ per batch (s)\end{tabular} & \begin{tabular}[c]{@{}c@{}}Avarage \\Testing Time \\ per batch (s)\end{tabular} & \begin{tabular}[c]{@{}c@{}}Average \\Number of \\Model\end{tabular}  \\ 
\midrule
\multirow{6}{*}{\begin{tabular}[c]{@{}c@{}}WeScatterNet with \\Regularization\\\textbf{using 25 percent}\\of labeled data\\LARGE SETTING\end{tabular}}

& Higgs     & 63.60& \textbf{19.57} & \textbf{17.51} & 2.51  \\
& Hepmass   & 83.44& \textbf{40.59} &\textbf{7.44}  & 1  \\
& Susy      & 75.67& \textbf{32.07}  &\textbf{11.27}  &  2 \\
& RLCPS     & 99.64& \textbf{52.66} & \textbf{4.79} & 1\\
& KDDCup    & 99.53& \textbf{65.53} & \textbf{9.01} & 1  \\
& PokerHand & 50.13& \textbf{8.59} & \textbf{8.33}  & 1  \\

\hline
\multirow{6}{*}{\begin{tabular}[c]{@{}c@{}}DEVFNN \\LARGE SETTING\end{tabular}}  
& Higgs$20\%$     & \textbf{65.28}& 6157.3  & N/A & 1 \\
& Hepmass$20\%$   & \textbf{84.12}& 5387.9   & N/A & 1 \\
& Susy$10\%$      & \textbf{78.96}& 2901.8& N/A & 1 \\
& RLCPS$10\%$     & \textbf{99.97}& 1730.6   & N/A  & 1 \\
& KDDCup$10\%$    &\textbf{99.79}& 5766.1  & N/A  & 1    \\
& PokerHand$50\%$ & 50.08& 2945.4   & N/A  & 1    \\ 
\hline
\multirow{6}{*}{\begin{tabular}[c]{@{}c@{}}pENsemble+ \\LARGE SETTING\end{tabular}} 
& Higgs$20\%$     & 47.05  & 7476.6  & 170.96  &    1  \\
& Hepmass$20\%$   & 80.73 & 2984.5   & 255.37  &    1  \\
& Susy$10\%$      & 76.96 &  3178.4  & 185.93  &    1  \\
& RLCPS$10\%$     & 99.8 & 1342.2   & 167.05  &    1  \\
& KDDCup$10\%$    & 99.77 & 253.41   & 168.24  &    1  \\
& PokerHand$50\%$ & 50.12 & 1062.8   & 219.03  &     1 \\
\hline

\multirow{6}{*}{\begin{tabular}[c]{@{}c@{}}DSSCN \\LARGE SETTING\end{tabular}} 
& Higgs$20\%$     & 65.27  &    213.72& 45.52  &    1  \\
& Hepmass$20\%$    &  60.41 & 371.25   &  54.30 & 1.82     \\
& Susy$10\%$       & 78.85  & 183.71   & 47.71  &  1    \\
& RLCPS$10\%$      & 99.90  & 135.49   & 47.37  &   1   \\
& KDDCup$10\%$     &  99.77 &  206.19  & 49.74  & 1      \\
& PokerHand$50\%$  & \textbf{50.35}& 558.46   & 69.52  &  1    \\
\hline

\multirow{6}{*}{\begin{tabular}[c]{@{}c@{}}pEnsemble \\LARGE SETTING\end{tabular}} 
& Higgs$20\%$     & 62.53  &  926.40  & 41.81  & 1     \\
& Hepmass$20\%$    & 81.09  & 983.08   & 47.84  & 1     \\
& Susy$10\%$       & 67.11  & 374.60   & 30.41  &  1    \\
& RLCPS$10\%$      & 99.98   & 210.53    &  29.53  &  1    \\
& KDDCup$10\%$     & 99.79  &  1316.87  & 34.06  &   1   \\
& PokerHand$50\%$  & 33.42  &  799.09  & 33.50  & 1     \\

\hline
\multirow{6}{*}{\begin{tabular}[c]{@{}c@{}}WeScatterNet with \\Regularization\\\textbf{using 25 percent}\\of labeled data\\SMALL SETTING\end{tabular}}    
& Higgs     &63.26 &6.43&5.1&2.01\\
& Hepmass   &83.45 &9.24 &2.748& 1 \\
& Susy      &75.7 &7.01 &2.25&  1 \\
& RLCPS     &99.64& 9.19 &1.95 & 1 \\
& KDDCup    & 99.41 &13.6 &3.23&  1 \\
& PokerHand &50.11& 4.67 &3.07 &  1 \\
\hline

\hline
\multirow{6}{*}{\begin{tabular}[c]{@{}c@{}}DEVFNN \\SMALL SETTING\end{tabular}}  
& Higgs$20\%$     & 65.21& 499.96   & N/A  & 1  \\
& Hepmass$20\%$   & 84.1& 441.55   & N/A   & 1 \\
& Susy$10\%$      & \textbf{78.92}& 405.09  & N/A  & 1 \\
& RLCPS$10\%$     & \textbf{99.91}& 182.89   & N/A  & 1    \\
& KDDCup$10\%$    & \textbf{99.78}& 712.65   & N/A  & 1    \\
& PokerHand$50\%$ & 50.11& 407.55   & N/A  & 1    \\ 
\hline
\multirow{6}{*}{\begin{tabular}[c]{@{}c@{}}pENsemble+ \\SMALL SETTING\end{tabular}} 
& Higgs$20\%$     &  47.07 & 2677.1   & 96.41  &    1.97  \\
& Hepmass$20\%$   & 82.88 & 554.52   & 115.63  &    1  \\
& Susy$10\%$      & 78.39 & 711.76   & 112.83  &    1.11  \\
& RLCPS$10\%$     & 99.91 & 116.23   &  120.12 &    1  \\
& KDDCup$10\%$    & 99.63 & 64.46   & 100.95  &    1  \\
& PokerHand$50\%$ & 50.10 & 962.49   &  107.4 &     1 \\

\hline

\multirow{6}{*}{\begin{tabular}[c]{@{}c@{}}DSSCN \\SMALL SETTING\end{tabular}} 
& Higgs$20\%$     & \textbf{65.22}  & 64.72    & 15.63  &    1  \\
& Hepmass$20\%$    & \textbf{84.15}  &  67.06  & 15.61  &    1  \\
& Susy$10\%$       & 78.81  & 53.74   &  14.81 &  1    \\
& RLCPS$10\%$      &  99.87 & 45.07   & 16.19  &  1    \\
& KDDCup$10\%$     &99.74   & 57.02   & 15.73  &  1    \\
& PokerHand$50\%$  & \textbf{50.84}  & 126.28   & 25.43  & 1     \\
\hline

\multirow{6}{*}{\begin{tabular}[c]{@{}c@{}}pEnsemble \\SMALL SETTING\end{tabular}} 
& Higgs$20\%$     & 63.47  & 494.15 & 16.07  & 1     \\
& Hepmass$20\%$    & 82.24  & 305.21   & 15.04  &  1    \\
& Susy$10\%$       & 73.92    & 53.59  &   13.22 & 1  \\
& RLCPS$10\%$      & 99.90  & 68.01   & 14.93  & 1     \\
& KDDCup$10\$$     & 99.71  & 540.56   & 14.74  & 1     \\
& PokerHand$50\%$  & 46.76  & 213.71   & 15.01  & 1     \\

\bottomrule
\end{tabular}
}
}
    \centering{}
  \end{center}

  \label{tab:NumResult_SingleNode}%
\end{table}

\subsection{Numerical Results}
Table 3 exhibits numerical results of all consolidated algorithms where they are run in two settings: large setting and small setting. The first one is directed to test the performance of an algorithm to handle large data streams while having low number of streams whereas the second one is arranged to evaluate the algorithm's performance in handling small data streams while having high number of data streams. WeScatterNet exploits only $25\%$ of labelled samples while other two algorithms utilize fully labelled training samples. 

Referring to Table 3, WeScatterNet's performance is competitive against its counterparts, ScattarNet and ScalablePANFIS in realm of large data stream. It outperforms other algorithms in Hepmass and Pokerhand while its performance is comparable to other two algorithms in the other two datasets. Note that WeScatterNet only utilizes 25\% label compared to other algorithms guided by fully labelled samples. This finding clearly shows that $DA^3$ method functions properly in preventing performance's drop due to the scarcity of labelled samples. On the other side, WeScatterNet beats other algorithms in four datasets in the small stream case: Higgs, Hepmass, Susy, PokerHand. Its performance in the other two datasets are comparable to other two algorithms. In addition to $DA^3$ method, this result demonstrates the advantage of data-free model fusion since it is carried out frequently in realm of small data streams, meanwhile, it still produces stable results. In the context of runtime, WeScatterNet is slower than ScalablePANFIS but enjoys semi-supervised environments rather than full supervision. It is caused by additional learning modules of WeScatterNet retarding its execution time. The execution time of WeScatterNet remains competitive compared to that ScatterNet. It is also observed that WeScatterNet and ScatterNet are implemented under an ensemble configuration while Scalable PANFIS works in the single node setting. Hence, Scalable PANFIS is faster than the two algorithms but its accuracy is worse than the two algorithms.  

\begin{figure}[htbp]
	\begin{centering}
		\includegraphics[width=10cm]{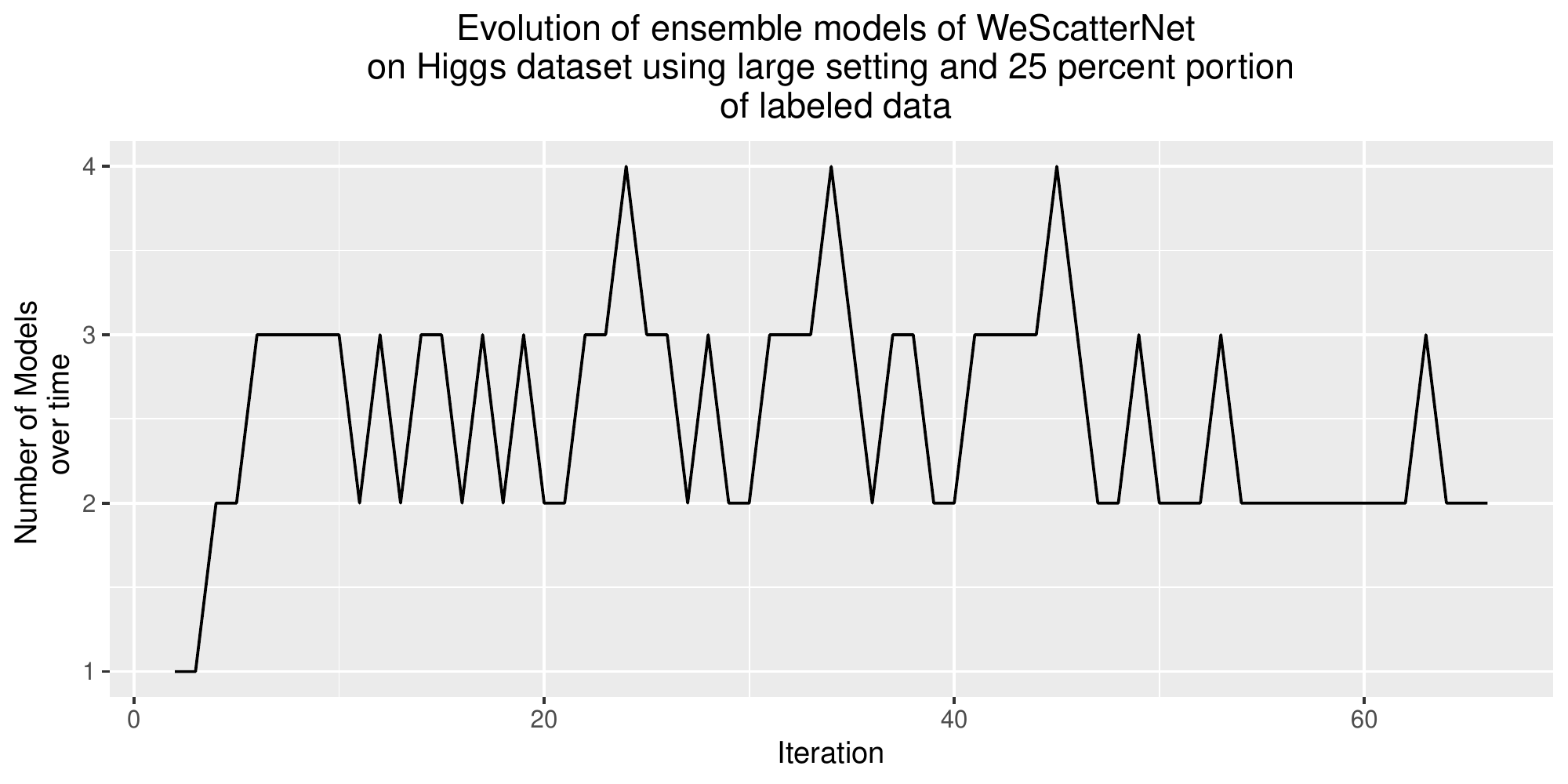}
		\par\end{centering}
	\caption{The evolution of ensemble structure: this figure depicts the efficacy of the ensemble pruning mechanism and the dynamic penalty and reward mechanism. The two mechanisms are capable of correcting the false drift alarms signalled by the drift detection mechanism. When the third and the fourth base learners are integrated into the ensemble structure, they are immediately discarded because they possess low voting weight. A low voting weight is caused by the penalty mechanism as a result of insufficient coverage of data samples. The number of base learners are consistent at two base learners. }
	\label{ModelEvol}
\end{figure}

\begin{figure}[htbp]
	\begin{centering}
		\includegraphics[width=10cm]{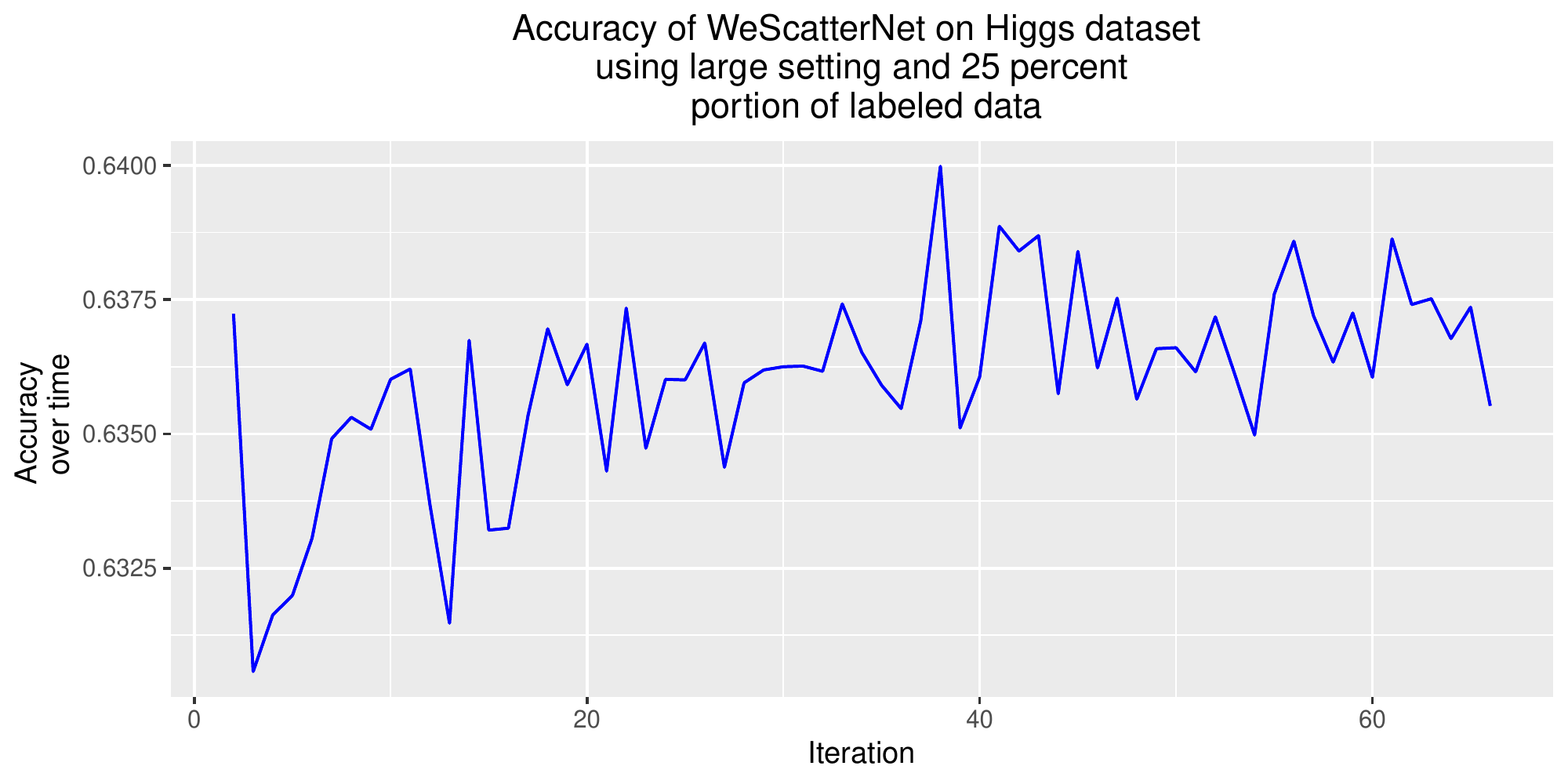}
		\par\end{centering}
	\caption{The trace of classification rates: it is illustrated that the classification rate shows an increasing trend. Furthermore, the standard deviation also decreases as the increase of observations. The structural learning mechanism does not harm the ensemble generalization performance as the classification rates are relatively stable in a small range.}
	\label{AccEvol}
\end{figure}

An open structure characteristic of WeScatterNet is visualized in Fig. \ref{ModelEvol} where a base classifier is flexibly added and removed in accordance with their performance during the training process. This figure also portrays the advantage of the base learning pruning strategy and the dynamic penalty and reward mechanism to identify an inactive base learner. The two mechanisms are capable of correcting false alarms to insert a new base learner triggered by the drift detection mechanism. The addition of the third and fourth base classifiers are responded by directly pruning these base classifiers since they possess a low voting weight. A low voting weight is resulted from the penalty mechanism lowering the influence of a base classifier in the weighted voting mechanism because they do not sufficiently cover data samples. In other words, a newly added base learner does not represent the majority of data points. The number of base learner converges to two base learners ultimately. This aspect is confirmed by Fig. \ref{AccEvol} where WeScatterNet's accuracy in overall shows an increasing trend. That is, the classification rates improves as the increase of observation while the standard deviation of the classification rates decreases. One can also observe that pruning the third and fourth base classifiers affect little to the classification performance of WeScatterNet. It is also portrayed in Fig. \ref{AccEvol} that the trace of classification rate hovers around a small range - less than 1\%. The false alarm and small fluctuation of classification rates are related to the drift-free characteristic of the Higgs dataset. Fig. \ref{ModelEvol} and Fig. \ref{AccEvol} refer to the WeScatterNet's performance in the Higgs dataset.

\begin{figure}[htbp]
	\begin{centering}
		\includegraphics[width=10cm]{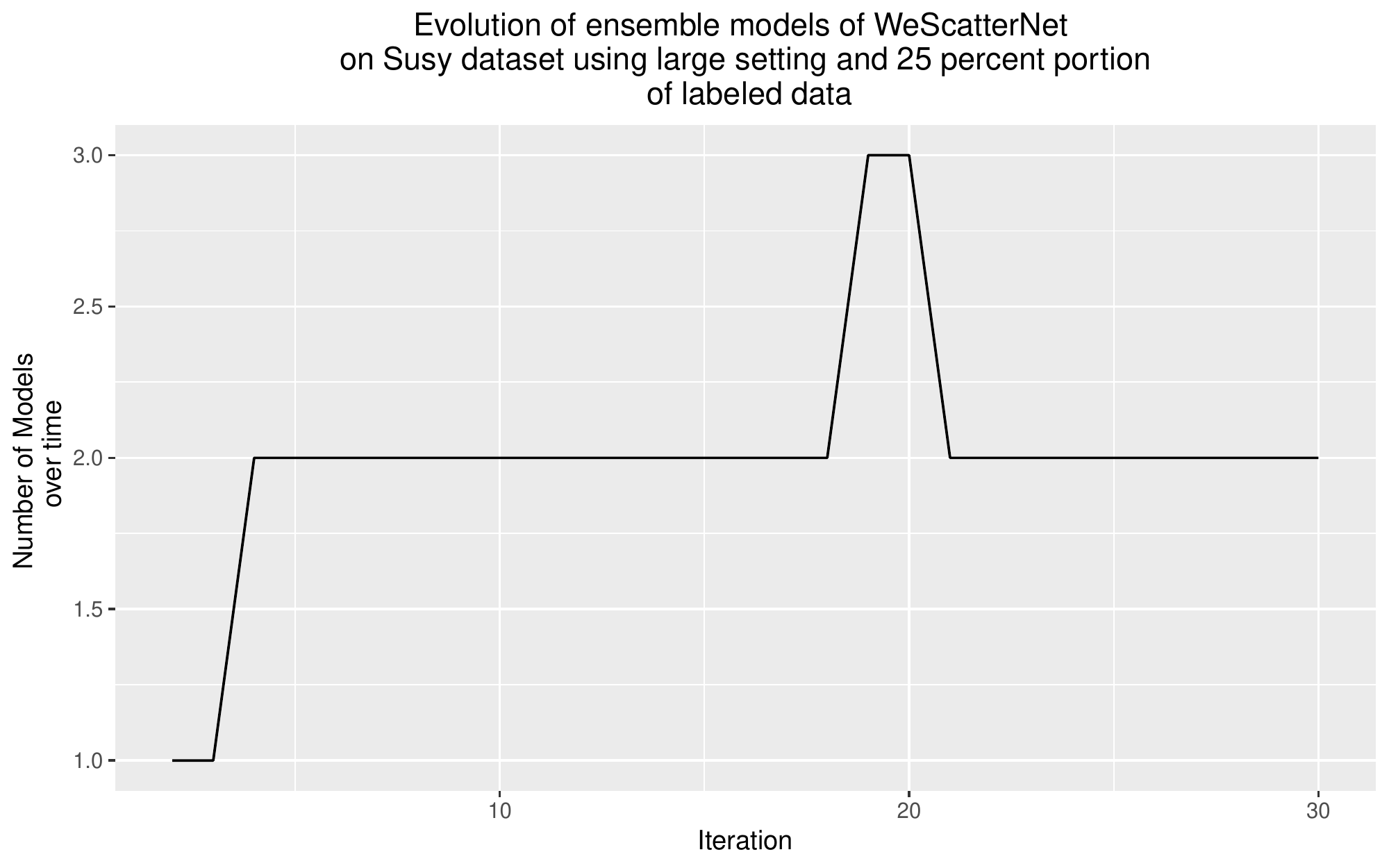}
		\par\end{centering}
	\caption{The evolution of ensemble structure: it is portrayed that the ensemble structure is relatively stable with 2 base classifiers. The introduction of the third base classifier occurs around $18^{th}$ data batch. This base learner does not help model's generalization and is thus pruned to improve the compactness of ensemble structure.}
	\label{SusyModelEvol}
\end{figure}

\begin{figure}[htbp]
	\begin{centering}
		\includegraphics[width=10cm]{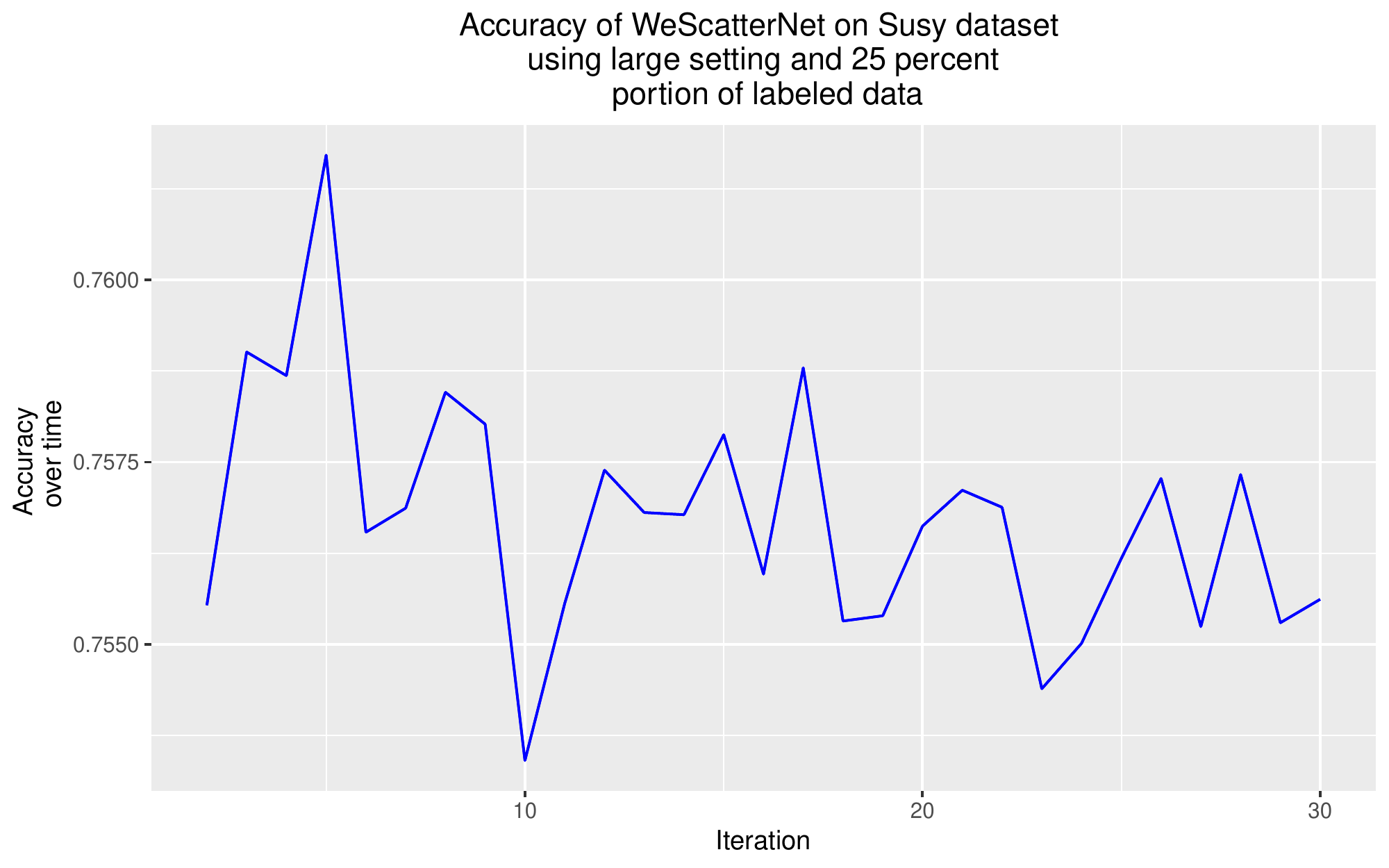}
		\par\end{centering}
	\caption{The trace of classification rates: it is observed that the classification rates of WeScatterNet fluctuate in a small range. The addition of the third base classifier contributes little to classifier's generalization. Pruning an inactive classifier does not undermine the classification rate of WeScatterNet.}
	\label{SusyAccEvol}
\end{figure}

Fig.\ref{SusyModelEvol} exhibits the evolution of WeScatterNet ensemble structure in the susy dataset. It differs from Fig. \ref{ModelEvol} where the ensemble structure is stable at two base learners. The third base learner is incorporated at the $18^{th}$ data batch but an extra base learner does not improve the model's generalization as indicated in Fig. \ref{SusyAccEvol}. One base learner is pruned where the ensemble structure converges at two base learners afterward. The pruning process is caused by an outdated base learner which does not cover the current concept well because it has a low voting weight. The trace of WeScatterNet's classification rates in the susy dataset is depicted in Fig. \ref{SusyAccEvol}. It is perceived that the classification rates of WeScatterNet fluctuates in a small range. This observation supports the local nature of the ensemble structure where the structural learning process does not harm the model's generalization. Small fluctuation of classification rates can be also linked to the fact that the susy dataset is drift-free. 

\subsection{Additional Comparisons}
The performance of WeScatterNet is also compared against four prominent algorithms: pENsemble+ \cite{pENsemble+}, DSSCN \cite{DSSCN}, DEVFNN \cite{DEVFNN} and pENsemble \cite{PratamaPedryczLughofer18}. The four algorithms are implemented under a single node setting. This comparison is to study whether WeScatterNet's performance remains competitive against popular fully-supervised data stream algorithms under the single node configuration. Nevertheless, the four algorithms are run through partial datasets because of their slow execution times as a result of the single node implementation. As with previous section, comparison is performed for both large setting and small setting while numerical results are produced by executing their published codes. Numerical results are reported in Table 4.

In realm of accuracy, DEVFNN and DSSCN produce the most encouraging result in both large and small settings but are not statistically significant to WeScatterNet only benefiting from $25\%$ labelled samples. Note that DEVFNN, DSSCN, pENsemble, pENsemble+ are executed only for a fraction of the overall datasets because of their execution times and in the single-node setting. WeScatterNet demonstrates obvious advantage over other four algorithms in the context of execution time where its training time and testing time are significantly faster than the four algorithms,i.e., its execution times are at least 10 times faster than other algorithms. It is worth mentioning that the success of a distributed algorithm can be declared if it delivers comparable accuracy to single-node algorithms while enjoying much faster execution time than those in the single node. WeScatterNet benefits from the distributed computing strategy of Apache spark for both training and testing processes where both processes can be executed in parallel across a number of worker nodes. By extension, the feasibility of WeScatterNet in the semi-supervised learning condition is also substantiated here because WeScatterNet deliver competitive accuracy compared to fully supervised algorithms with only $25\%$ labelled samples.


\begin{table}[htbp]
  \centering
  \caption{Numerical Results of WeScatterNet on six datasets using different label proportions for both large and small settings: it is observed that performance of our algorithm is stable across different label proportions.}
  \scalebox{0.6}{
 {\renewcommand{\arraystretch}{0.8} 

    \begin{tabular}{cccccccccc}
    \toprule
    \multicolumn{1}{c}{\multirow{2}[4]{*}{Dataset}} & \multicolumn{1}{c}{\multirow{2}[4]{*}{\begin{tabular}[c]{@{}c@{}} Proportion of \\labeled data\end{tabular}}} & \multicolumn{4}{c}{Performance (average per batch)} & 
    \multicolumn{1}{c}{\multirow{2}[4]{*}{\begin{tabular}[c]{@{}c@{}} NPseu \\(per batch)\end{tabular}}} &
    \multicolumn{1}{c}{\multirow{2}[4]{*}{\begin{tabular}[c]{@{}c@{}} NLabel \\(per batch)\end{tabular}}} &
    \multicolumn{1}{c}{\multirow{2}[4]{*}{\begin{tabular}[c]{@{}c@{}} NAug \\(per batch)\end{tabular}}} &
    \multicolumn{1}{c}{\multirow{2}[4]{*}{\begin{tabular}[c]{@{}c@{}} NSampleBatch \\(per batch)\end{tabular}}} \\
\cmidrule{3-6}          &       & Accuracy (\%) & TrTime (s) & TsTime (s) & Nmodels &       &       &       &  \\
    \midrule
    \multicolumn{1}{c}{\multirow{4}[2]{*}{\begin{tabular}[c]{@{}c@{}} Higgs \\LARGE SETTING\end{tabular}}} 
          & 0.1   &63.27$\pm$0.32&15$\pm$1.63&16.02$\pm$4.49&2.32$\pm$0.66&(55.8$\pm$4.6)K&16.7K&16.7K&166.7K\\
          & 0.25  &63.60$\pm$0.18&19.57$\pm$1.42&17.51$\pm$4.37&2.51$\pm$0.64&(52.3$\pm$2.5)K&41.7K&41.7K&166.7K\\
          & 0.5  &63.69$\pm$0.15&24.45$\pm$1.77&16.74$\pm$4.65&2.42$\pm$0.66&(36.2$\pm$1.2)K&83.3K&83.3K&166.7K\\
          & 0.75  &63.66$\pm$0.15&32.68$\pm$2.17&28.25$\pm$10.02&3.78$\pm$1.39&(19$\pm$1)K&125K&125K&166.7K\\
    \midrule
   \multicolumn{1}{c}{\multirow{4}[2]{*}{\begin{tabular}[c]{@{}c@{}}Hepmass \\LARGE SETTING\end{tabular}}} 
            & 0.1   &83.46$\pm$0.1&33.16$\pm$2.38&7.35$\pm$0.36&1$\pm$0&(105.6$\pm$4.1)K&16.7K&16.7K&166.7K\\
          & 0.25  &83.44$\pm$0.09&40.59$\pm$1.83&7.44$\pm$0.38&1$\pm$0&(96.6$\pm$1.2)K&41.7K&41.7K&166.7K\\
          & 0.5   &83.48$\pm$0.1&46.85$\pm$1.96&7.92$\pm$0.47&1$\pm$0&(66.9$\pm$0.3)K&83.3K&83.3K&166.7K\\
          & 0.75  &83.58$\pm$0.09&52.95$\pm$3.31&8$\pm$0.44&1$\pm$0&(33.9$\pm$0.1)K&125K&125K&166.7K\\
    \midrule
   \multicolumn{1}{c}{\multirow{4}[2]{*}{\begin{tabular}[c]{@{}c@{}}Susy \\LARGE SETTING\end{tabular}}} 
            & 0.1   &75.8$\pm$0.15&25.03$\pm$3.57&6.03$\pm$0.26&1$\pm$0&(91.5$\pm$7.1)K&16.7K&16.7K&166.7K\\
          & 0.25 &75.67$\pm$0.17&32.07$\pm$2.21&11.27$\pm$2.08&2$\pm$0.38&(87$\pm$2.3)K&41.7K&41.7K&166.7K\\
          & 0.5   &75.59$\pm$0.17&36.6$\pm$2.24&10.52$\pm$2.42&1.79$\pm$0.41&(60.3$\pm$0.9)K&83.3K&83.3K&166.7K\\
          & 0.75 &75.35$\pm$0.15&41.05$\pm$2.36&6.37$\pm$0.21&1$\pm$0&(30.8$\pm$0.3)K&125K&125K&166.7K\\
    \midrule
      \multicolumn{1}{c}{\multirow{4}[2]{*}{\begin{tabular}[c]{@{}c@{}}RLCPS \\LARGE SETTING\end{tabular}}} 
      
         & 0.1   &99.64$\pm$0.01&51.31$\pm$1.42&4.73$\pm$0.17&1$\pm$0&(150$\pm$0.1)K&16.7K&16.7K&166.7K\\
          & 0.25  &99.64$\pm$0.01&52.32$\pm$1.21&4.77$\pm$0.19&1$\pm$0&(124.9$\pm$0.2)K&41.7K&41.7K&166.7K\\
          & 0.5   &99.64$\pm$0.01&53.93$\pm$1.93&4.87$\pm$0.32&1$\pm$0&(83.1$\pm$0.1)K&83.3K&83.3K&166.7K\\
          & 0.75  &99.64$\pm$0.01&55.58$\pm$2.19&5.16$\pm$0.28&1$\pm$0&(41.7$\pm$0)K&125K&125K&166.7K\\
    \midrule
     \multicolumn{1}{c}{\multirow{4}[2]{*}{\begin{tabular}[c]{@{}c@{}}KDDCup \\LARGE SETTING\end{tabular}}} 
        & 0.1   &99.54$\pm$0.06&62.67$\pm$2.69&9.82$\pm$0.48&1$\pm$0&(150.9$\pm$0.2)K&16.9K&16.9K&168.9K\\
          & 0.25 &99.53$\pm$0.06&64.04$\pm$2.88&9.11$\pm$0.58&1$\pm$0&(125.9$\pm$0.1)K&42.2K&42.2K&168.9K\\
          & 0.5  &99.68$\pm$0.03&69.04$\pm$3.45&16.68$\pm$3.39&1.82$\pm$0.39&(84$\pm$0.1)K&84.5K&84.5K&168.9K\\
          & 0.75  &99.6$\pm$0.06&73.86$\pm$4.3&10.26$\pm$0.57&1$\pm$0&(42$\pm$0)K&126.7K&126.7K&168.9K\\
    \midrule
   \multicolumn{1}{c}{\multirow{4}[2]{*}{\begin{tabular}[c]{@{}c@{}}Poker Hand \\LARGE SETTING\end{tabular}}} 
      & 0.1  &50.13$\pm$0.09&5.3$\pm$1.26&8.69$\pm$0.22&1$\pm$0&(0$\pm$0)K&17.1K&17.1K&170.8K\\
          & 0.25 &50.13$\pm$0.09&9.12$\pm$2.76&9.66$\pm$0.32&1$\pm$0&(0$\pm$0)K&42.7K&42.7K&170.8K\\
          & 0.5  &50.13$\pm$0.09&13.04$\pm$4.35&7.89$\pm$0.25&1$\pm$0&(0$\pm$0)K&85.4K&85.4K&170.8K\\
          & 0.75  &50.13$\pm$0.09&20.2$\pm$7.74&9.26$\pm$0.1&1$\pm$0&(0$\pm$0)K&128.1K&128.1K&170.8K\\
    \midrule
    \midrule
   \multicolumn{1}{c}{\multirow{4}[2]{*}{\begin{tabular}[c]{@{}c@{}}Higgs \\SMALL SETTING\end{tabular}}} 
      & 0.1   &62.5$\pm$0.42&4.75$\pm$0.4&5.5$\pm$1.63&2.11$\pm$0.63&(19.3$\pm$1.2)K&5.6K&5.6K&55.6K\\
          & 0.25  &63.26$\pm$0.35&6.43$\pm$0.4&5.1$\pm$0.6&2.01$\pm$0.24&(18.6$\pm$1)K&13.9K&13.9K&55.6K\\
          & 0.5   &63.47$\pm$0.24&8.72$\pm$0.52&5.67$\pm$0.77&1.99$\pm$0.25&(12.8$\pm$0.5)K&27.8K&27.8K&55.6K\\
          & 0.75 &63.59$\pm$0.28&10.23$\pm$0.79&6.14$\pm$1.4&2.07$\pm$0.39&(6.3$\pm$0.2)K&41.7K&41.7K&55.6K\\
    \midrule
   \multicolumn{1}{c}{\multirow{4}[2]{*}{\begin{tabular}[c]{@{}c@{}}Hepmass \\SMALL SETTING\end{tabular}}} 
      & 0.1   &83.36$\pm$0.18&7.41$\pm$0.4&5.17$\pm$0.33&1.99$\pm$0.1&(35$\pm$1.1)K&5.6K&5.6K&55.6K\\
          & 0.25 &83.45$\pm$0.16&9.24$\pm$0.42&2.74$\pm$0.13&1$\pm$0&(31.9$\pm$0.3)K&13.9K&13.9K&55.6K\\
          & 0.5  &83.48$\pm$0.16&11.32$\pm$0.56&2.91$\pm$0.18&1$\pm$0&(22.3$\pm$0.1)K&27.8K&27.8K&55.6K\\
          & 0.75  &83.49$\pm$0.16&12.73$\pm$0.57&5.79$\pm$0.41&1.99$\pm$0.1&(11.3$\pm$0)K&41.7K&41.7K&55.6K\\
    \midrule
 \multicolumn{1}{c}{\multirow{4}[2]{*}{\begin{tabular}[c]{@{}c@{}}Susy \\SMALL SETTING\end{tabular}}} 
    & 0.1  &75.7$\pm$0.23&5.57$\pm$0.42&2.24$\pm$0.15&1$\pm$0&(29.6$\pm$1.8)K&5.6K&5.6K&55.6K\\
          & 0.25  &75.7$\pm$0.22&7.01$\pm$0.42&2.25$\pm$0.13&1$\pm$0&(28.4$\pm$0.6)K&13.9K&13.9K&55.6K\\
          & 0.5  &75.29$\pm$0.22&9.38$\pm$0.52&2.39$\pm$0.13&1$\pm$0&(20.3$\pm$0.2)K&27.8K&27.8K&55.6K\\
          & 0.75  &75.24$\pm$0.27&10.25$\pm$0.6&4.63$\pm$0.54&1.98$\pm$0.21&(10.2$\pm$0.1)K&41.7K&41.7K&55.6K\\
    \midrule
  \multicolumn{1}{c}{\multirow{4}[2]{*}{\begin{tabular}[c]{@{}c@{}}RLCPS \\SMALL SETTING\end{tabular}}} 
     & 0.1  &99.64$\pm$0.02&8.88$\pm$0.44&1.83$\pm$0.13&1$\pm$0&(50$\pm$0.1)K&5.6K&5.6K&55.6K\\
          & 0.25  &99.64$\pm$0.02&9.19$\pm$0.58&1.89$\pm$0.55&1$\pm$0&(41.7$\pm$0)K&13.9K&13.9K&55.6K\\
          & 0.5   &99.64$\pm$0.02&10.32$\pm$0.52&2.07$\pm$0.15&1$\pm$0&(27.8$\pm$0)K&27.8K&27.8K&55.6K\\
          & 0.75  &99.64$\pm$0.02&10.5$\pm$0.55&1.92$\pm$0.1&1$\pm$0&(13.9$\pm$0)K&41.7K&41.7K&55.6K\\
    \midrule
  \multicolumn{1}{c}{\multirow{4}[2]{*}{\begin{tabular}[c]{@{}c@{}}KDDCup \\SMALL SETTING\end{tabular}}} 
     & 0.1   &99.18$\pm$0.06&13.21$\pm$0.63&3.75$\pm$0.16&1$\pm$0&(50.3$\pm$0)K&5.6K&5.6K&56.3K\\
          & 0.25  &99.41$\pm$0.05&13.97$\pm$0.75&3.68$\pm$0.27&1$\pm$0&(41.9$\pm$0)K&14.1K&14.1K&56.3K\\
          & 0.5   &99.5$\pm$0.05&15.51$\pm$0.94&3.44$\pm$0.2&1$\pm$0&(28$\pm$0)K&28.2K&28.2K&56.3K\\
          & 0.75  &99.58$\pm$0.04&16.53$\pm$1.05&3.55$\pm$0.23&1$\pm$0&(14$\pm$0)K&42.2K&42.2K&56.3K\\
    \midrule
 \multicolumn{1}{c}{\multirow{4}[2]{*}{\begin{tabular}[c]{@{}c@{}}Poker Hand \\SMALL SETTING\end{tabular}}}
    & 0.1  &50.13$\pm$0.17&3.66$\pm$0.34&3.44$\pm$0.15&1$\pm$0&(0$\pm$0)K&5.7K&5.7K&56.9K\\
          & 0.25  &50.11$\pm$0.17&4.78$\pm$0.71&3.39$\pm$0.19&1$\pm$0&(0$\pm$0)K&14.2K&14.2K&56.9K\\
          & 0.5  &50.13$\pm$0.17&7.66$\pm$1.06&3.48$\pm$0.18&1$\pm$0&(0$\pm$0)K&28.5K&28.5K&56.9K\\
          & 0.75  &50.11$\pm$0.17&9.68$\pm$1.61&3.4$\pm$0.15&1$\pm$0&(0$\pm$0)K&42.7K&42.7K&56.9K\\
    \midrule
    
      \multicolumn{10}{c}{{\begin{tabular}[c]{@{}c@{}}\textbf{TrTime}: Training Time; \textbf{TsTime}: Testing Time; \textbf{Nmodels}: Number of models/base classifiers; \\
      \textbf{NPseu}: Average number of pseudolable processed in each batch\\ \textbf{NLabel}: Number of labeled data; \textbf{NAug}: Number of augmented label; \\ \textbf{NSampleBatch}: Number of samples processed in each batch (in fully supervised learning condition)\end{tabular}}} \\

    \bottomrule
    \end{tabular}%

 }
}
  \label{tab:NumResultsWeScatterNetReg}%
\end{table}%

\subsection{Effect of Labelled Samples}
This section analyzes different quantities of labelled samples to the performance of WeScatterNet tested in two settings: small stream and large stream. WeScatterNet's performance is evaluated with four different label proportions: 10\%, 25\%, 50\%, 75\% where the numerical results are offered in Table \ref{tab:NumResultsWeScatterNetReg}. 

It is perceived that different numbers of labels do not vary the performance of WeScatterNet significantly. The increase of labelled samples only provides minor effect on the performance of WeScatterNet. It is seen that the maximum difference in performance between 10\% and 75\% labels is less than 0.75\%. This fact demonstrates the advantage of WeScatterNet in achieving decent performance despite low label quantity. In other words, $DA^3$ algorithm is capable of compensating the loss of generalization power as a result of low labelled samples. 

It is observed that the increase of labelled samples slows down the training time of WeScatterNet. This finding is reasonable because it means the increase of data instances to be processed in the base learner training process. On the other side, the increase of label proportion automatically decreases the number of pseudo label. Note that the self-labelling mechanism to generate pseudo-labels focuses only on unlabelled samples. It is also perceived that the number of pseudo labels is significant in which some of them might be noisy. The noisy pseudo labels are handled using the regularization method to circumvent the loss of generalization. The generation of augmented labels is also illustrated here in which it is produced by injecting a controlled noise to labelled samples without changing its label. As a result, the number of augmented label is akin to the number of original label.

\begin{table}[htbp]
  \centering
  \caption{Ablation Study : Numerical Results of WeScatterNet on three conditional settings using 25 percent of labeled data. Each learning module contributes positively to the performance of our algorithm. 
  }
  \scalebox{0.6}{
 {\renewcommand{\arraystretch}{0.8} 

    \begin{tabular}{cccccccccc}
    \toprule
    \multicolumn{1}{c}{\multirow{2}[4]{*}{Dataset}} & \multicolumn{1}{c}{\multirow{2}[4]{*}{\begin{tabular}[c]{@{}c@{}} Conditional \\Settings\end{tabular}}} & \multicolumn{4}{c}{Performance (average per batch)} & 
    \multicolumn{1}{c}{\multirow{2}[4]{*}{\begin{tabular}[c]{@{}c@{}} NPseu \\(per batch)\end{tabular}}} &
    \multicolumn{1}{c}{\multirow{2}[4]{*}{\begin{tabular}[c]{@{}c@{}} NLabel \\(per batch)\end{tabular}}} &
    \multicolumn{1}{c}{\multirow{2}[4]{*}{\begin{tabular}[c]{@{}c@{}} NAug \\(per batch)\end{tabular}}} &
    \multicolumn{1}{c}{\multirow{2}[4]{*}{\begin{tabular}[c]{@{}c@{}} NSampleBatch \\(per batch)\end{tabular}}} \\
\cmidrule{3-6}          &       & Accuracy (\%) & TrTime (s) & TsTime (s) & Nmodels &       &       &       &  \\
    \midrule
   
    \multicolumn{1}{c}{\multirow{4}[2]{*}{\begin{tabular}[c]{@{}c@{}} Higgs \\LARGE SETTING\end{tabular}}} 
          & \textbf{Reg}  &63.6$\pm$0.18&19.57$\pm$1.42&17.51$\pm$4.37&2.51$\pm$0.64&(52.3$\pm$2.5)K&41.7K&41.7K&166.7K\\
          & \textbf{NoReg} &55.42$\pm$3.42&46.95$\pm$14.69&16.18$\pm$5.22&2.34$\pm$0.8&(106.4$\pm$27.2)K&41.7K&41.7K&166.7K\\
          & \textbf{NoAug} &63.54$\pm$0.23&14.75$\pm$1.01&14.33$\pm$4.02&2.15$\pm$0.62&(48.3$\pm$2.1)K&41.7K&0&166.7K\\
    \midrule
   \multicolumn{1}{c}{\multirow{4}[2]{*}{\begin{tabular}[c]{@{}c@{}}Higgs \\SMALL SETTING\end{tabular}}} 
              & \textbf{Reg}
              &63.26$\pm$0.35&6.43$\pm$0.4&5.1$\pm$0.6&2.01$\pm$0.24&(18.6$\pm$1)K&13.9K&13.9K&55.6K\\
          & \textbf{NoReg}  &53.36$\pm$1.23&11.06$\pm$0.74&5.01$\pm$0.52&1.99$\pm$0.19&(40.9$\pm$2.5)K&13.9K&13.9K&55.6K\\
          & \textbf{NoAug}    &63.01$\pm$0.29&5.03$\pm$0.33&5.52$\pm$0.71&2.04$\pm$0.26&(17.9$\pm$1.5)K&13.9K&0&55.6K\\
    \midrule
   \multicolumn{1}{c}{\multirow{4}[2]{*}{\begin{tabular}[c]{@{}c@{}}Hepmass \\LARGE SETTING\end{tabular}}} 
         & \textbf{Reg} &83.44$\pm$0.09&39.64$\pm$1.65&7.12$\pm$0.31&1$\pm$0&(96.6$\pm$1.2)K&41.7K&41.7K&166.7K\\
         & \textbf{NoReg} &83.39$\pm$0.1&32.38$\pm$2.62&13.79$\pm$1.53&1.98$\pm$0.22&(83.9$\pm$3.8)K&41.7K&41.7K&166.7K\\
          & \textbf{NoAug}   &83.48$\pm$0.09&33.87$\pm$1.72&7.16$\pm$0.28&1$\pm$0&(92.1$\pm$2.4)K&41.7K&0&166.7K\\
    \midrule
      \multicolumn{1}{c}{\multirow{4}[2]{*}{\begin{tabular}[c]{@{}c@{}}Hepmass \\SMALL SETTING\end{tabular}}} 
      
     & \textbf{Reg}  
     &83.45$\pm$0.16&8.63$\pm$0.36&2.58$\pm$0.1&1$\pm$0&(31.9$\pm$0.3)K&13.9K&13.9K&55.6K\\
         & \textbf{NoReg}   &81.87$\pm$1.94&7.46$\pm$0.4&5.1$\pm$0.31&1.99$\pm$0.1&(27.2$\pm$0.8)K&13.9K&13.9K&55.6K\\
          & \textbf{NoAug}  &83.29$\pm$0.19&7.35$\pm$0.34&5.06$\pm$0.28&1.99$\pm$0.1&(30.3$\pm$0.7)K&13.9K&0&55.6K\\  
    \midrule

    \multicolumn{10}{c}{{\begin{tabular}[c]{@{}c@{}}\textbf{Reg}: WeScatterNet using regularization; \textbf{NoReg}: WeScatterNet without using regularization; \\\textbf{NoAug}: WeScatterNet using regularization in the absence of augmented label\end{tabular}}} \\

    \bottomrule
    \end{tabular}%

 }
}
  \label{tab:Ablation}%
\end{table}%

\subsection{Ablation Study}
This section studies the effect of each learning modules to the resultant performance. WeScatterNet is simulated under three learning configurations: A. WeScatterNet is set with the absence of regularization method; B. WeScatterNet is arranged with the absence of augmented label, C. WeScatterNet is run in the centralized learning setting in the single node. Table \ref{tab:Ablation} sums up the performance of WeScatterNet in two learning configurations, whereas Table \ref{tab:AblationSingle} shows the performance of WeScatterNet in two different environmental settings (single and distributed). Our ablation study is performed by exploiting $25\%$ labelled samples while two simulation conditions, namely small stream and large stream, are explored. Our numerical study is performed in the Higgs, Hepmass and Susy problems.

\begin{table}[htbp]
  \centering
  \caption{Ablation Study : Numerical Results of WeScatterNet for both single node and distributed nodes (with Regularization using 25 percent of labeled data). It is demonstrated that there exist subtle differences between single node and distributed node.}
  
  \scalebox{0.6}{
 {\renewcommand{\arraystretch}{0.8} 

    \begin{tabular}{cccccccccc}
    \toprule
    \multicolumn{1}{c}{\multirow{2}[4]{*}{Dataset}} & \multicolumn{1}{c}{\multirow{2}[4]{*}{\begin{tabular}[c]{@{}c@{}} Environment \\Settings\end{tabular}}} & \multicolumn{4}{c}{Performance (average per batch)} & 
    \multicolumn{1}{c}{\multirow{2}[4]{*}{\begin{tabular}[c]{@{}c@{}} NPseu \\(per batch)\end{tabular}}} &
    \multicolumn{1}{c}{\multirow{2}[4]{*}{\begin{tabular}[c]{@{}c@{}} NLabel \\(per batch)\end{tabular}}} &
    \multicolumn{1}{c}{\multirow{2}[4]{*}{\begin{tabular}[c]{@{}c@{}} NAug \\(per batch)\end{tabular}}} &
    \multicolumn{1}{c}{\multirow{2}[4]{*}{\begin{tabular}[c]{@{}c@{}} NSampleBatch \\(per batch)\end{tabular}}} \\
\cmidrule{3-6}          &       & Accuracy (\%) & TrTime (s) & TsTime (s) & Nmodels &       &       &       &  \\
    \midrule
   
    \multicolumn{1}{c}{\multirow{2}[2]{*}{\begin{tabular}[c]{@{}c@{}} Higgs \\LARGE SETTING\end{tabular}}} 
          & \textbf{Single Node}  &63.78$\pm$0.14&2253.07$\pm$71.69&151.06$\pm$5.38&1$\pm$0&(48$\pm$1.8)K&41.7K&41.7K&166.7K\\
          & \textbf{Distributed} &63.6$\pm$0.18&19.57$\pm$1.42&17.51$\pm$4.37&2.51$\pm$0.64&(52.3$\pm$2.5)K&41.7K&41.7K&166.7K\\
        
    \midrule
   \multicolumn{1}{c}{\multirow{2}[2]{*}{\begin{tabular}[c]{@{}c@{}}Higgs \\SMALL SETTING\end{tabular}}} 
          & \textbf{Single Node}  &63.78$\pm$0.22&265.4$\pm$12.8&29.54$\pm$0.85&1$\pm$0&(15.9$\pm$0.6)K&13.9K&13.9K&55.6K\\
          & \textbf{Distributed}   &63.26$\pm$0.35&6.43$\pm$0.4&5.1$\pm$0.6&2.01$\pm$0.24&(18.6$\pm$1)K&13.9K&13.9K&55.6K\\
    \midrule
   \multicolumn{1}{c}{\multirow{2}[2]{*}{\begin{tabular}[c]{@{}c@{}}Hepmass \\LARGE SETTING\end{tabular}}} 
          & \textbf{Single Node}  
          &83.47$\pm$0.1&2430.17$\pm$78.75&148.43$\pm$5.29&1$\pm$0&(96.5$\pm$1.1)K&41.7K&41.7K&166.7K\\
          & \textbf{Distributed} &83.44$\pm$0.09&39.64$\pm$1.65&7.12$\pm$0.31&1$\pm$0&(96.6$\pm$1.2)K&41.7K&41.7K&166.7K\\
    \midrule
      \multicolumn{1}{c}{\multirow{2}[2]{*}{\begin{tabular}[c]{@{}c@{}}Hepmass \\SMALL SETTING\end{tabular}}} 
      
         & \textbf{Single Node}  &83.46$\pm$0.17&303.72$\pm$14.57&27.7$\pm$1.13&1$\pm$0&(32$\pm$0.3)K&13.9K&13.9K&55.6K\\
         & \textbf{Distributed}  &83.45$\pm$0.16&8.63$\pm$0.36&2.58$\pm$0.1&1$\pm$0&(31.9$\pm$0.3)K&13.9K&13.9K&55.6K\\
    \midrule
   \multicolumn{1}{c}{\multirow{2}[2]{*}{\begin{tabular}[c]{@{}c@{}}Susy \\LARGE SETTING\end{tabular}}} 
          & \textbf{Single Node} &75.85$\pm$0.12&1415.22$\pm$110.07&151.01$\pm$8.78&1$\pm$0&(83.2$\pm$2.3)K&41.7K&41.7K&166.7K\\
          & \textbf{Distributed} &75.67$\pm$0.17&32.07$\pm$2.21&11.27$\pm$2.08&2$\pm$0.38&(87$\pm$2.3)K&41.7K&41.7K&166.7K\\
    \midrule
      \multicolumn{1}{c}{\multirow{2}[2]{*}{\begin{tabular}[c]{@{}c@{}}Susy \\SMALL SETTING\end{tabular}}} 
      
         & \textbf{Single Node}  &75.91$\pm$0.19&176.73$\pm$11.05&27.77$\pm$0.97&1$\pm$0&(27.3$\pm$0.5)K&13.9K&13.9K&55.6K\\
         & \textbf{Distributed}  &75.7$\pm$0.22&7.01$\pm$0.42&2.25$\pm$0.13&1$\pm$0&(28.4$\pm$0.6)K&13.9K&13.9K&55.6K\\

    \bottomrule
    \end{tabular}%

 }
}
  \label{tab:AblationSingle}%
\end{table}%

It is seen that the performance of WeScatterNet suffers from the absence of regularization approach where $3-8\%$ performance degradation is observed if the regularization method is deactivated. This fact is understood from the auto-regularization mechanism plays key role in protecting WeScatterNet from performance's degradation due to noisy pseudo label. The use of augmented label also contributes positively to the performance of WeScatterNet. The absence of such mechanism brings the performance of WeScatterNet down as a result of the consistency regularization. That is, a sample is varied by injecting controlled perturbation without changing its class label. Furthermore, WeScatterNet performs comparably to its single node variant where almost identical performance is resulted. This case demonstrates the advantage of the model fusion strategy where it does not cause any performance's compromise. On the other hand, the advantage of distributed processing for data streams is clearly demonstrated where its run time significantly reduces. 
The advantage of distributed computing strategy is clearly demonstrated in Table \ref{tab:AblationSingle} in which it expedites the execution time by over 200 times. This observation is more evident in the large setting than in the small setting in which the size of data stream is large thus confirming the scalability of distributed computing strategy of Apache spark. The single node configuration affects little to the number of pseudo label and augmented label since the generation of pseudo label and augmented label are done in the centralized fashion.

\section{Conclusion}
This paper presents Weakly Supervised Scalable Teacher Forcing Network (WeScatterNet) as a solution of large-scale data streams under semi-supervised mode where only small fractions of data samples in streams are annotated. WeScatterNet is developed under a distributed computing platform of Apache Spark making possible for parallel execution of large streams in an efficient manner. This implementation is underpinned by a data-free model fusion method performing model compression after the parallel computing stage. The problem of partially labelled data instances (semi-supervised learning) is addressed by the $DA^3$ method performing the label enrichment mechanism followed by the dynamic regularization step to overcome noisy pseudo labels. The issue of non-stationary distribution is overcome by having the global and local drift handling mechanisms where the global drift handling mechanism controls the growing and pruning processes of base learners while the local drift handling mechanism is carried out by the growing and pruning processes of fuzzy rules of a base learner thereby actualizing a fully dynamic network structure in both ensemble level and the base learner level. The base learner of WeScatterNet is built upon a simplified TS fuzzy model where both rule premise and rule conclusion share the same parameter, i.e., a linear hyperplane enabled by the hyperplane clustering concept. The teacher forcing principle is adopted to address the dependence on the target variable when performing inferences. 

The advantage of WeScatterNet has been numerically validated in six large-scale data stream problems with only $25\%$ labelled samples. It is demonstrated that WeScatterNet delivers highly competitive accuracy in both large setting and small setting even compared to fully supervised algorithms and those running in the single computing node environments. Our numerical study also investigates different proportions of labelled samples, 10\%,25\%,50\%,75\% in which surprisingly these variations do not affect the performance of WeScatterNet. As analyzed in the ablation study, the regularization principle plays significant role in underpinning the performance of WeScatterNet while the use of augmented label slightly improves the accuracy. Another important finding is in the subtle difference in performance between the distributed performance of WeScatterNet and the single node performance of WeScatterNet. This fact signifies the success of distributed implementation of Apache spark including the model compression stage. Our future work is directed to study the multistream classification problem. That is, a model is supposed to handle many data streams running simultaneously. This issue not only demands algorithm's competence in handling data streams but also in performing domain adaptation thereby leading to a domain-invariant network.

\section{Acknowledgement}
This work is supported by Ministry of Education Republic of Singapore Tier 1 research grant. The second author acknowledges that this work was supported by Pawsey Supercomputing Centre through the use of advanced computing resources. The third author acknowledges the support by the 'LCM --- K2 Center for Symbiotic Mechatronics' within the framework of the Austrian COMET-K2 program.

.

\bibliography{mybibfile}

\end{document}